\documentclass[sigconf,natbib=true]{acmart}
\usepackage{multirow}
\usepackage{subcaption}
\usepackage{color}
\usepackage{enumitem}
\usepackage{xcolor}
\usepackage{graphicx}
\usepackage{subcaption} % 用于子图功能
\usepackage[table]{xcolor}
\definecolor{mygray}{RGB}{230,230,230}
\usepackage[linesnumbered,ruled,vlined]{algorithm2e}

\usepackage{bbm}
\usepackage{mathtools}

\setlist[itemize]{leftmargin=*}

\AtBeginDocument{%
  \providecommand\BibTeX{{%
    \normalfont B\kern-0.5em{\scshape i\kern-0.25em b}\kern-0.8em\TeX}}}

\setcopyright{acmlicensed}
\copyrightyear{2018}
\acmYear{2018}
\acmDOI{XXXXXXX.XXXXXXX}

\acmConference[Conference acronym 'XX]{Make sure to enter the correct
  conference title from your rights confirmation emai}{June 03--05,
  2018}{Woodstock, NY}
\acmBooktitle{Woodstock '18: ACM Symposium on Neural Gaze Detection,
 June 03--05, 2018, Woodstock, NY} 
\acmISBN{978-1-4503-XXXX-X/18/06}

\begin{document}

%%
%% The "title" command has an optional parameter,
%% allowing the author to define a "short title" to be used in page headers.
% \title{Not Everything Needs to be Forgotten:\\Recommendation Unlearning via Guided Filtering}
% \title{Forgetting the Outdated: Efficient Recommendation Unlearning via Guided Filtering}
% \title{Retaining Beneficial Knowledge:\\Recommendation Unlearning via Guided Filtering}
% \title{Retaining Makes Better:\\Recommendation Unlearning via Guided Filtering}
% \title{Preserving Makes Better:\\Recommendation Unlearning via Guided Filtering}
% \title{Filtering User Preferences from Recommendation Unlearning}
% \title{An Efficient Guided Filtering Framework to Erase Outdated Preferences for Recommendation Unlearning}
\title{Tail-Aware Data Augmentation for Long-Tail Sequential Recommendation}

%%
%% The "author" command and its associated commands are used to define
%% the authors and their affiliations.
%% Of note is the shared affiliation of the first two authors, and the
%% "authornote" and "authornotemark" commands
%% used to denote shared contribution to the research.
\author{Yizhou Dang}
\affiliation{%
  \institution{Software College, Northeastern University}
  \city{Shenyang}
  \country{China}
}
\email{dangyz@mails.neu.edu.cn}

\author{Zhifu Wei}
\affiliation{%
  \institution{School of Computer Science and Engineering, Northeastern University}
  \city{Shenyang}
  \country{China}
}
\email{weizf@mails.neu.edu.cn}

\author{Minhan Huang}
\affiliation{%
  \institution{Software College, Northeastern University}
  \city{Shenyang}
  \country{China}
}
\email{huangminhan@stumail.neu.edu.cn}

\author{Lianbo Ma}
\affiliation{%
  \institution{Software College, Northeastern University}
  \city{Shenyang}
  \country{China}
}
\email{malb@swc.neu.edu.cn}

\author{Jianzhe Zhao}
\affiliation{%
  \institution{Software College, Northeastern University}
  \city{Shenyang}
  \country{China}
}
\email{zhaojz@swc.neu.edu.cn}

\author{Guibing Guo}
\affiliation{%
  \institution{Software College, Northeastern University}
  \city{Shenyang}
  \country{China}
}
\email{guogb@swc.neu.edu.cn}

\author{Xingwei Wang}
\affiliation{%
  \institution{School of Computer Science and Engineering, Northeastern University}
  \city{Shenyang}
  \country{China}
}
\email{wangxw@mail.neu.edu.cn}

%%
%% By default, the full list of authors will be used in the page
%% headers. Often, this list is too long, and will overlap
%% other information printed in the page headers. This command allows
%% the author to define a more concise list
%% of authors' names for this purpose.
\renewcommand{\shortauthors}{Yizhou and Zhifu, et al.}

%%
%% The abstract is a short summary of the work to be presented in the
%% article.
\begin{abstract}
Sequential recommendation (SR) learns user preferences based on their historical interaction sequences and provides personalized suggestions. In real-world scenarios, most users can only interact with a handful of items, while the majority of items are seldom consumed. This pervasive long-tail challenge limits the model's ability to learn user preferences. Despite previous efforts to enrich tail items/users with knowledge from head parts or improve tail learning through additional contextual information, they still face the following issues: 1) They struggle to improve the situation where interactions of tail users/items are scarce, leading to incomplete preferences learning for the tail parts. 2) Existing methods often degrade overall or head parts performance when improving accuracy for tail users/items, thereby harming the user experience. In this work, we propose \textbf{T}ail-\textbf{A}ware \textbf{D}ata \textbf{A}ugmentation (TADA) for long-tail sequential recommendation, which enhances the interaction frequency for tail items/users while maintaining head performance, thereby promoting the model's learning capabilities for the tail. Specifically, we first capture the co-occurrence and correlation among low-popularity items by a linear model. Building upon this, we design two tail-aware augmentation operators, T-Substitute and T-Insert. The former replaces the head item with a relevant item, while the latter utilizes co-occurrence relationships to extend the original sequence by incorporating both head and tail items. The augmented and original sequences are mixed at the representation level to preserve preference knowledge. We further extend the mix operation across different tail-user sequences and augmented sequences to generate richer augmented samples, thereby improving tail performance. Comprehensive experiments demonstrate the superiority of our method. The codes are provided at \url{https://github.com/KingGugu/TADA}.

\end{abstract}

%%
%% The code below is generated by the tool at http://dl.acm.org/ccs.cfm.
%% Please copy and paste the code instead of the example below.
%%


%%
%% Keywords. The author(s) should pick words that accurately describe
%% the work being presented. Separate the keywords with commas.

%% A "teaser" image appears between the author and affiliation
%% information and the body of the document, and typically spans the
%% page.
% \begin{teaserfigure}
%   \includegraphics[width=\textwidth]{sampleteaser}
%   \caption{Seattle Mariners at Spring Training, 2010.}
%   \Description{Enjoying the baseball game from the third-base
%   seats. Ichiro Suzuki preparing to bat.}
%   \label{fig:teaser}
% \end{teaserfigure}

% \received{20 February 2007}
% \received[revised]{12 March 2009}
% \received[accepted]{5 June 2009}

%%
%% This command processes the author and affiliation and title
%% information and builds the first part of the formatted document.
\maketitle

\section{Introduction}

\begin{figure}[!t]
    \centering

    \begin{subfigure}[t]{0.45\textwidth}
        \centering
        \includegraphics[width=\textwidth]{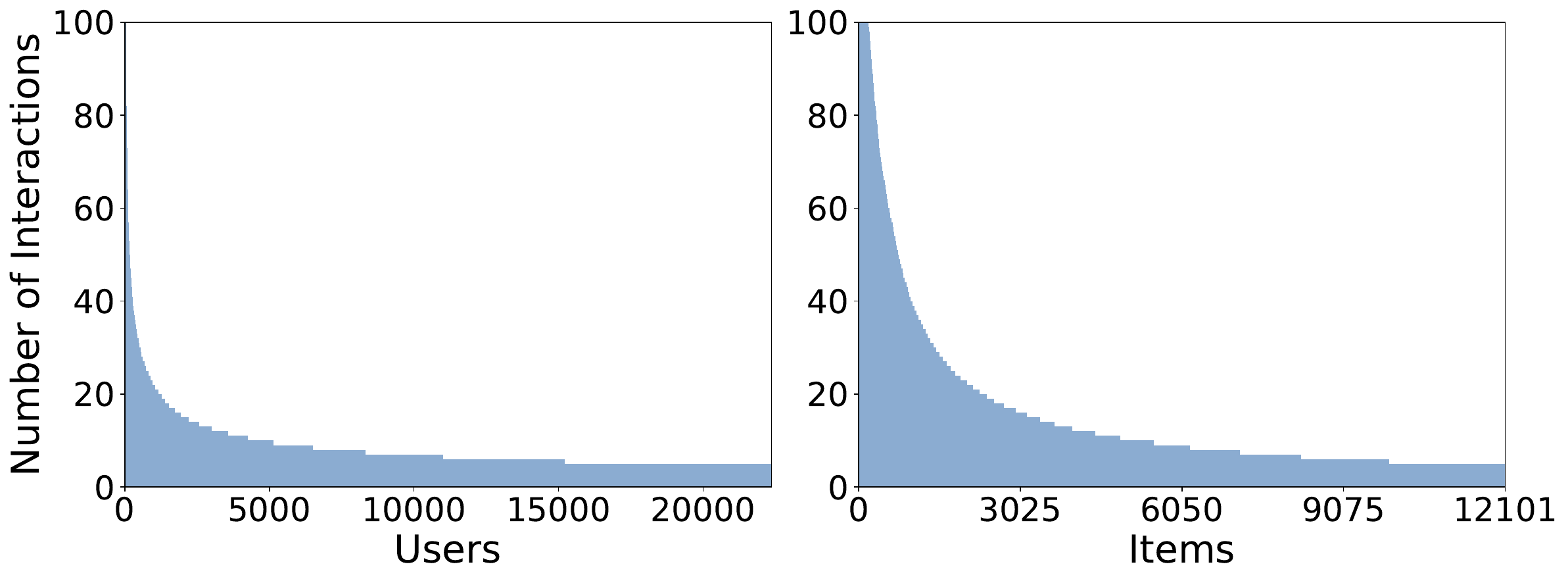}
        \caption{Visualization of long-tail users (left) and items (right) in Beauty dataset. Existing knowledge transfer-based methods fail to effectively increase the number of interactions for the tail.}
        \label{fig:sub1}
    \end{subfigure}

    \vspace{4.5pt} 

    \begin{subfigure}[b]{0.45\textwidth}
        \centering
        \includegraphics[width=\textwidth]{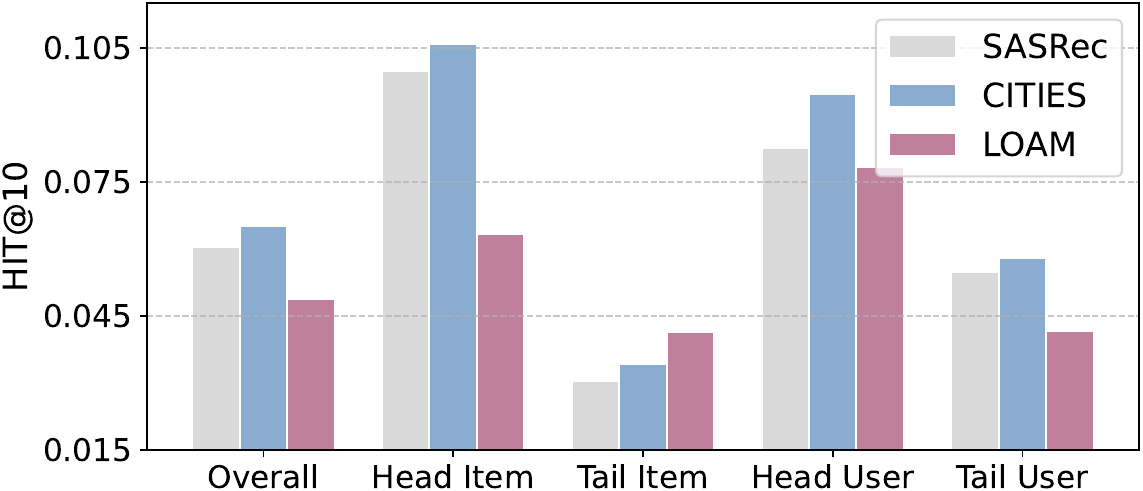}
        \caption{Existing works may result in overall and head performance damage (i.e. seesaw effect) while improving tail performance.}
        \label{fig:sub2}
    \end{subfigure}
    \vspace{-5pt}
    \caption{The (a) long tail issue and (b) seesaw effect.}
    \vspace{-5pt}
    \label{fig:1-Eeample}
\end{figure}

As an indispensable component of web services, sequential recommendation has garnered widespread attention due to its high degree of consistency with real-world scenarios \cite{kang2018self,zhai2025combinatorial,di2025personalized,di2025pifgsr}. It models user preferences based on user-item interaction sequences and recommends items that users are likely to interact with next. Numerous SR models based on various architectures have been proposed, achieving significant advancements \cite{hidasi2015session,he2016fusing,kang2018self,zhou2022filter}. Despite their effectiveness, the widespread long-tail problem limits the practicality of the model \cite{jang2020cities,liu2024llm}. Specifically, due to users' limited purchasing power and interaction capacity, most users can only interact with a tiny portion of items on the platform \cite{dang2024data}. This results in short historical sequences for these users (i.e., long-tail users), making it difficult for models to learn accurate user behavior patterns from them \cite{kim2023melt}. From the perspective of items, only a handful of popular items are viewed or purchased frequently by users, while most items are rarely consumed (i.e., long-tail items), making it challenging for models to learn accurate item representations \cite{liu2024llm}. As shown in Figure \ref{fig:1-Eeample} (a), the real-world Beauty dataset demonstrates a high proportion of both long-tail users and items.

To address the long-tail problem, the primary solution in existing research is knowledge transfer \cite{jang2020cities,kim2023melt,yang2023loam}. The representative methods, CITIES \cite{jang2020cities}, leveraged representations learned from popular items to enrich representations of tail items, thereby compensating for the poor convergence of embeddings caused by limited interactions. On the other hand, some efforts attempted to incorporate additional interactions for long-tail users, enhancing the model's ability to learn and reason about their preferences \cite{jiang2021sequential,liu2023diffusion}. For example, ASReP \cite{jiang2021sequential} employs a reversely pre-trained transformer to generate pseudo-prior items for short sequences. Then, fine-tune the pre-trained transformer to predict the following item.

Despite the progress, we two unresolved issues remain: \textbf{(1) Interaction Sparsity:} Existing methods struggle to increase interaction frequency for long-tail users and items fundamentally. Sparse interactions constrain the model's ability to learn user preferences. Mainstream SR models treat item representations as the primary learnable parameters, while user representations are usually modeled implicitly and embedded in the model's output representation of the interaction sequence \cite{kang2018self}. Consequently, the number of item interactions directly determines the convergence of item embeddings during training, thereby affecting prediction accuracy. Although representations from head items can partially refine tail item embeddings, interaction sparsity significantly constrains the model's learning capacity. It is worth emphasizing that items added by ASReP are mostly head items, thereby cannot improve the sparseness of tail items \cite{liu2024llm}. \textbf{(2) Seesaw Phenomenon:} While improving performance for tail-end users and items, existing methods may result in overall performance degradation and loss of top-tier performance (i.e., the seesaw problem), thereby compromising the overall user experience. As shown in Figure \ref{fig:1-Eeample} (b), although LOAM \cite{yang2023loam} improves the recommendation accuracy for tail users and items, it reduces performance of other parts. 

To tackle the above challenges, in this work, we propose \textbf{T}ail-\textbf{A}ware \textbf{D}ata \textbf{A}ugmentation (TADA) for long-tail sequential recommendation. Our core idea is to leverage crafted augmentation modules to increase interactions of tail items and users, which enhances long-tail performance while avoiding the loss of head and overall performance caused by knowledge transfer. Specifically, we first constructed a linear item-item Similarity model \cite{choi2025linear}, utilizing ridge regression to model co-occurrence relationships between items (particularly long-tail unpopular items). Based on this correlation, we designed two tail-aware data augmentation operators, T-Substitute and T-Insert. The T-Substitute prioritizes replacing head items in sequences with relevant long-tail items. T-Insert prioritizes inserting relevant head items before tail items to boost the model's prediction for those tail items. Subsequently, we mix the replaced sequences with their original sequences at the representation level, while mixing the inserted sequences with their extended original sequences, thereby generating novel yet semantically relevant samples \cite{dang2025augmenting}. Subsequently, we extended the mixing operation within the cross-sequence augmentation module to encompass different long-tail sequences and their augmented counterparts. It enables the model to optimize representations from diverse augment samples for both head and tail sequences in a differentiated manner, thereby achieving a consistent improvement in overall performance. Extensive experiments on various datasets and backbones to validate the superiority of our proposed method.

In summary, our work makes the following contributions:
\begin{itemize}
  \item We highlight the interaction sparsity and seesaw phenomenon of existing long-tail SR methods and propose a \textbf{T}ail-\textbf{A}ware \textbf{D}ata \textbf{A}ugmentation (TADA) method to enhance the interaction frequency for tail items/users while maintaining head performance.

  \item We design several key modules. Two tail-aware operators, T-Substitute and T-Insert, perform augmentation on the tail items and generate augmented samples through fusion at the representation level. Cross-sequence mixing further provides diverse samples by utilizing different original and augmented sequences.

  \item We conduct comprehensive experiments on real-world datasets with representative backbone SR models and different baselines to demonstrate the effectiveness and generalizability of TADA.
\end{itemize}

\section{Related Work}

\noindent \textbf{General SR.} Sequential Recommendation (SR) is a key task to predict the next accessed item based on users' interaction sequences \cite{zhao2023sequential,zhao2023cross,dang2025data}. Early works used Markov Chains \cite{he2016fusing} to model user behaviors, focusing on correlations between the next item and recent interactions. Later, CNNs \cite{yuan2019simple} and RNNs \cite{2022cmnrec} were adopted: GRU4Rec \cite{hidasi2015session} used GRU to model interest evolution, while Caser \cite{tang2018personalized} mapped items to temporal/latent spaces and learned patterns via convolutional filters. With the success of self-attention \cite{vaswani2017attention}, Transformer-based SR models emerged. SASRec \cite{kang2018self} used Transformer layers to learn item importance and characterize transition correlations. STOSA \cite{fan2022sequential} represented items as Gaussian distributions and designed Wasserstein Self-Attention for item relationships. Beyond Transformers, studies found filtering algorithms reduce noise \cite{zhou2022filter}, leading to the proposal of an all-MLP architecture with learnable filters. LRURec \cite{yue2024linear} designed a linear recurrence with matrix diagonalization to capture transition patterns and a recursive parallelization framework that accelerates training.

\vspace{0.3em}

\noindent \textbf{Long-Tail SR.} To mitigate the long-tail problem, existing research primarily focuses on transferring knowledge from head users and items to the tail to enrich representations of tail users and items \cite{kim2019sequential,jang2020cities,liu2025consistency}. The representative work, CITIES \cite{jang2020cities}, enhanced the quality of the tail-item embeddings by training
an embedding-inference function using multiple contextual head items. Following this idea, MELT \cite{kim2023melt} designed a mutually reinforced dual-branch framework to enhance performance for both tail users and items. MASR \cite{hu2022memory} proposed centroid-wise and cluster-wise memory banks to store historical samples. It searched similar sequences and copied information to enhance the tail items. The above ideas are also used in session-based recommendation \cite{liu2020long,yang2023loam}. With the rise of large language models (LLMs), some efforts have also explored leveraging textual semantic information and multimodal data to enrich and enhance long-tail users and items \cite{liu2024llm,li2025reembedding,li2025listwise}. For example, LLM-ESR \cite{liu2024llm} combined semantics from LLMs and collaborative signals from conventional SR for tail items, and proposed a retrieval augmented self-distillation method to enhance tail users. Unlike the above efforts, our proposed TADA with tail-aware augmentation modules are tailored from a data perspective, directly addressing the interaction sparsity of tail users and items.

\vspace{0.3em}

\noindent \textbf{Data Augmentation for SR.}
Given widespread data sparsity in SR, numerous data augmentation methods have been proposed with notable progress \cite{xie2022contrastive,dang2024data,di2025federated}, mainly categorized into heuristic-based and model-based approaches. Heuristic methods generate new samples by perturbing the original sequences. Early ones include Sliding Window \cite{tang2018personalized} (splitting sequences into sub-sequences) and Deletion \cite{tan2016improved}. With contrastive learning’s application in SR, operators like Crop, Mask, Reorder, Insert, and Substitute emerged \cite{liu2021contrastive,xie2022contrastive}, later improved by incorporating time intervals \cite{dang2023ticoserec}, user intents \cite{chen2022intent}, behavior periodicity \cite{tian2023periodicity}, and relevance/diversity \cite{dang2025augmenting}. Due to the low controllability of heuristic-based methods, model-based methods have emerged with learnable generative modules. DiffASR \cite{liu2023diffusion} adopted diffusion models with guidance for sequence generation. ASReP \cite{liu2021augmenting} used a reverse-pre-trained Transformer to generate pseudo-prior items for short sequences. A recent work, DR4SR \cite{yin2024dataset} proposed a data-level framework to refine sequences for better informativeness and generalization. However, these methods lack specialized augmentation strategies tailored for long-tail items and users, failing to increase interaction volumes at the tail end effectively. Consequently, the long-tail problem remains unresolved.

\section{Preliminaries}
\noindent \textbf{Problem Formulation of SR.} We use $\mathcal{U}$ and $\mathcal{V}$ to denote the user set and the item set, respectively. Each user $u \in \mathcal{U}$ is associated with a sequence of interacted items in chronological order $s_u=[v_1, \ldots, v_j, \ldots, v_{\left|s_u\right|}]$, where $v_j \in \mathcal{V}$ indicate the item that user $u$ has interacted with at time step $j$ and $\left|s_u\right|$ is the sequence length. Given the interaction sequence $s_u$, sequential recommendation aims to predict the most possible item $v^{*}$ that user $u$ will interact with at time step $\left|s_u\right|+1$, which can be formulated as:
\begin{equation}
    \underset{v^{*} \in \mathcal{V}}{\arg \max} \;\; P\left(v_{\left|s_u\right|+1}=v^{*} \mid s_u \right).
\end{equation}

\noindent \textbf{Head/Tail Splitting.} Following the previous on long-tail SR, we split the users and items into tail and head groups \cite{jang2020cities,liu2024llm}. Firstly, we sort the users and items by the sequence length $\left|s_u\right|$ and the popularity of the item $v$ (i.e., the total interaction number). Then, take out the top $20 \%$ users and items as head user and head item according to the Pareto principle \cite{box1986analysis}, denoted as $\mathcal{U}_{\text {head}}$ and $\mathcal{V}_{\text{head}}$. The rest of the users and items are the tail user and tail item, i.e., $\mathcal{U}_{\text {tail}}=\mathcal{U} - \mathcal{U}_{\text {head}}$ and $\mathcal{V}_{\text {tail}}=\mathcal{V} - \mathcal{V}_{\text {head}}$ \cite{liu2024llm}. For long-tail issues in SR, we aim to enhance the performance for $\mathcal{U}_{\text{tail}}$ and $\mathcal{V}_{\text{tail}}$.

\noindent \textbf{Embeddings and Representations.} The standard SR paradigm \cite{kang2018self, dang2025data} maintains an item embedding matrix $\mathbf{M}_\mathcal{V} \in \mathbb{R}^{|\mathcal{V}| \times D}$. The matrix projects the high-dimensional one-hot representation of an item to a low-dimensional, dense representation. Given an original user sequence $s_u=[v_1, v_2, \ldots, v_{\left|s_u\right|}]$, the embedding $\operatorname{Look-up}$ operation will be applied for $\mathbf{M}_\mathcal{V}$ to get a sequence of item representation, i.e., $E_u=[e_{v_1}, e_{v_2}, \ldots, e_{v_{\left|s_u\right|}}]$. We omit the padding for short sequences and the interception for long sequences. By feeding the sequence representation $E_u$  into the Encoder, we can obtain the output representation of the sequence $H_u$. In sequential recommendation, user representations are usually modeled implicitly, so $H_u$ can also be interpreted as user representations \cite{kang2018self}.

\section{Ours: TADA}

\begin{figure*}[!t]
  \centering
  \includegraphics[scale=0.37]{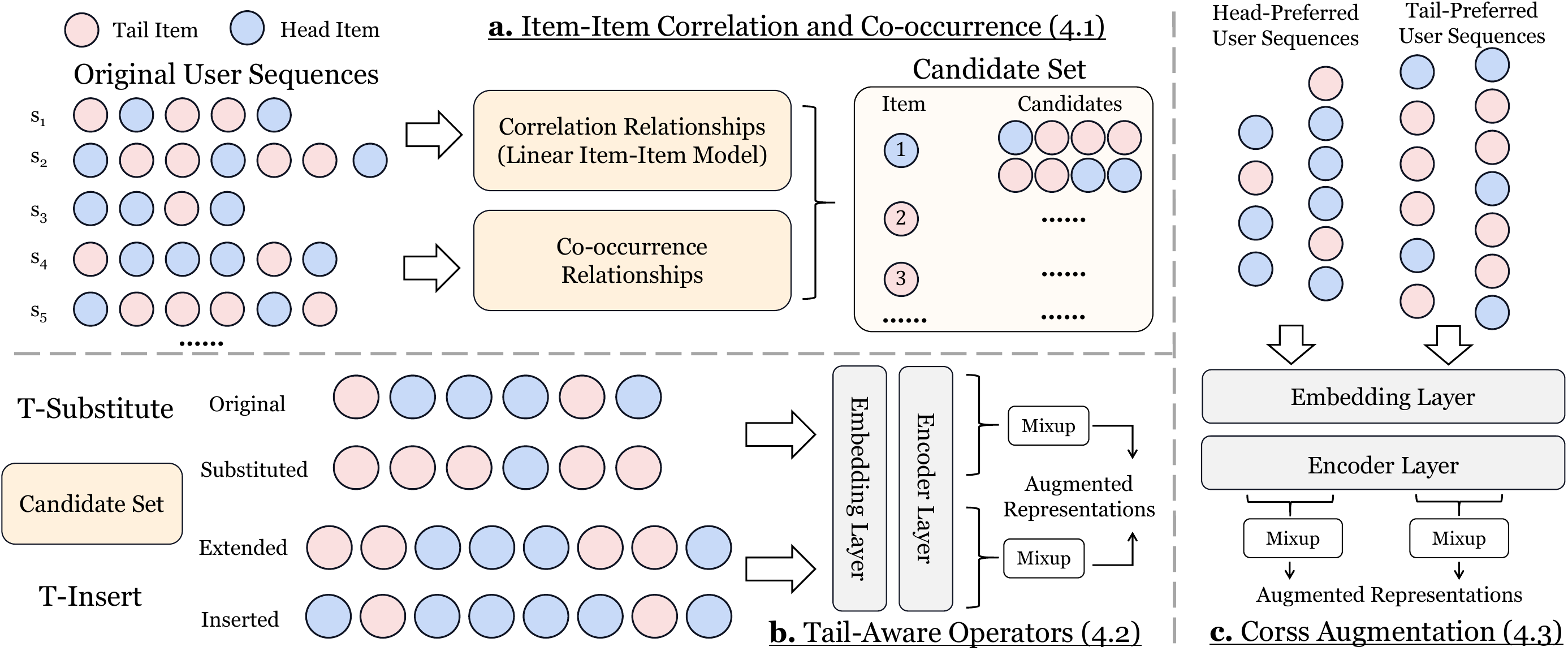}
  \vspace{-1em}
  \caption{The overall framework of our proposed TADA.}
  \label{fig:framework}
  \vspace{-1em}
\end{figure*}

The overall framework of our proposed TADA is illustrated in Figure \ref{fig:framework}. It first constructs an augmentation candidate set based on correlation and co-occurrence relationships. Subsequently, two tail-aware augmentation operators augment long-tail interactions based on this candidate set. cross augmentation further extends this augmentation across different tail and augmented sequences. All augmented samples are utilized for model training. The comprehensive overview of the notations is given in Table~\ref{tab:notions}.

\begin{table}[t]
\centering
\caption{Frequently used notations in this paper.}
\vspace{-1em}
\renewcommand\arraystretch{1.0}
\scalebox{0.85}{
\begin{tabular}{c|c}
\toprule \textbf{Notation} & \textbf{Description} \\
\midrule
$\mathcal{U},\mathcal{V},u,v$ & User set, item set, a user, an item\\
$\mathcal{U}_{head},\mathcal{U}_{tail}$ & Head user set, tail user set. \\
$\mathcal{V}_{head},\mathcal{V}_{tail}$ & Head item set, tail item set. \\
$s_u,|s_u|$ & Interaction sequence of user $u$, sequence length\\
$N,B$ & Maximum sequence length, batch size. \\
$M_V,M_D,M_S$ & Item embedding, interaction, and item-item matrices. \\
$\alpha$ & Hyperparameter for Beta distribution. \\
$\lambda$ & Mixup weight from Beta($\alpha, \alpha$). \\
$p$ & Probability of selecting items for augmentation. \\
$a, b$ & Hyperparameters for uniform distribution. \\
$cr_v,cc_v,c_v$ & Different candidate sets for item $v$. \\
$s_u{\prime},s_u^e$ & Augmented and extended sequence from operators. \\
$E_u,E_u{\prime}$ & Sequence of item embeddings. \\
$H_u,H_u{\prime}$ & Output representation from the encoder. \\
$H_u^{ao},H_u^{ac}$ & Final augmented representation. \\
$E_u^+/E_u^{ac+}, E_u^-/E_u^{ac-}$ & Embeddings of positive and negative items for $s_u$. \\
$\sigma(\cdot)$ & Sigmoid function in BCE loss. \\
$\beta$ & Threshold for classifying sequences. \\
\bottomrule
\end{tabular}}
\vspace{-1em}
\label{tab:notions}
\end{table}

\subsection{Item-Item Correlation and Co-occurrence}
To provide suitable candidate augmentation items for our operators, we construct candidate item sets based on item-item correlation relationships and co-occurrence relationships.

\vspace{0.3em}

\noindent \textbf{Correlation Relationships.} Previous work primarily relied on cosine similarity to search and select the similar candidate items, which not only resulted in $O(|\mathcal{V}|^2)$ computational complexity but also required performing calculations during each screening alongside model training \cite{liu2021contrastive}. Here, we employ a linear model to compute similarity scores in advance before model training begins \cite{choi2025linear}. Furthermore, some research has demonstrated that linear models can better capture co-occurrence among less popular items (long-tail items), aligning perfectly with our requirements.

The linear model takes the user-item interaction matrix $\mathbf{M}_\mathcal{D} \in \mathbb{R}^{|\mathcal{U}| \times |\mathcal{V}|}$ ($y_{u v}=1$ represents observed interactions in user sequences, and $y_{u v}=0$ represents unobserved interactions) as input and finally output an item-item similarity matrix $\mathbf{M}_\mathcal{S} \in \mathbb{R}^{|\mathcal{V}| \times |\mathcal{V}|}$. We set constraints for the diagonal elements to be less than or equal to $\xi \in[0,1)$ to avoid the trivial solution and to consider the case of repeated items \cite{choi2025linear}. The learning process can be formulated as:
\begin{equation}
\hat{\mathbf{M}_\mathcal{S}}=\underset{\mathbf{M}_\mathcal{S}}{\arg \min}\|\mathbf{M}_\mathcal{D}-\mathbf{M}_\mathcal{D} \cdot \mathbf{M}_\mathcal{S}\|_F^2+\lambda\|\mathbf{M}_\mathcal{S}\|_F^2 \text { s.t.} \operatorname{diag}(\mathbf{M}_\mathcal{S}) \leq \xi,
\label{eq:linear}
\end{equation}
where $\xi$ is the hyperparameter to control diagonal constraints, and $\lambda$ is the regularization coefficient. The closed-form solution of Equation (\ref{eq:linear}) can be formulate as \cite{choi2021session}:
\begin{equation}
\hat{\mathbf{M}_\mathcal{S}}=\mathbf{I}-\mathbf{P} \cdot \operatorname{diagMat}(\gamma), \quad \gamma_j= \begin{cases}\lambda & \text { if} 1-\lambda \mathbf{P}_{j j} \leq \xi \\ \frac{1-\xi}{\mathbf{P}_{j j}} & \text { otherwise}\end{cases}
\end{equation}
where $\mathbf{P}=\left(\mathbf{M}_\mathcal{D}^{\top} \mathbf{M}_\mathcal{D}+\lambda \mathbf{M}_\mathcal{D}\right)^{-1}$, $\operatorname{diagMat}(\cdot)$ denotes the diagonal matrix, and $\gamma \in \mathbb{R}^n$ is the vector for checking the diagonal constraint of $\hat{\mathbf{M}}$. The $\hat{\mathbf{M}_\mathcal{S}}$ captures the similarity between items. For each item $v$, we selected Top-$K$ similar items as augmentation candidates from correlation relationships, denoted as $cr_{v}=\{v_{c1}, v_{c2}, \ldots, v_{K}\}$

\vspace{0.3em}

\noindent \textbf{Co-occurrence Relationships.} Since correlation relationships can only account for similarity at the representation level, they cannot model the relative positional information of items within a sequence. Therefore, we also consider the first-order co-occurrence relationships between different items, specifically the sequential relationships before and after each item in the interaction sequence. For each head item, we select the tail items from all its preceding and following items across all sequences as its co-occurrence set. For each tail item, we select all its preceding items across all sequences as its co-occurrence set. This co-occurrence set can be regarded as the set of behaviors that lead to the occurrence of the tail item. For each item $v$, the co-occurrence candidate set is denoted as $cc_{v}=\{v_{c1}, v_{c2}, \ldots, v_{|cc_{v}|}\}$. We take the union of $cr_{v}$ and $cc_{v}$ as the final candidate set for each item, denoted as $c_{v}$:
\begin{equation}
    c_{v} = cr_{v} \cup cc_{v}
\end{equation}

\subsection{Tail-Aware Operators}
Existing augmentation methods randomly select augmentation positions and items during enhancement, failing to effectively increase interactions for tail-end users and items \cite{liu2021contrastive,liu2021augmenting,xie2022contrastive}. Therefore, the model's performance on the tail remains unsatisfactory. To tackle this issue, We propose two tail-aware data augmentation operators, T-Substitute and T-insert, which effectively increase the interaction frequency of tail items and users to promote model learning. 

\vspace{0.3em}

\noindent \textbf{T-Substitute.} Given an original sequence $s_u=[v_1, \ldots, v_j, \ldots, v_{\left|s_u\right|}]$, this operator iterates through all items in the sequence and selects each head item with probability $p$. The indices of these selected items will be recorded as  $\left\{\operatorname{idx}_1, \operatorname{idx}_2, \ldots, \operatorname{idx}_i\right\}$. Unlike traditional operators that perform augmentation with a fixed $p$, T-Substitute extends the possibilities by drawing $p$ from a uniform distribution:
\begin{equation}
rate \sim \operatorname{Uniform}(a, b),
\label{eq:rate}
\end{equation}
where $a$ and $b$ are hyper-parameters and $0$ \textless $a$ \textless $b$ \textless $1$. The changing probability values can provide more novel augmented samples during the model training process, improving the performance and robustness of the model. Then, we replace each selected head item $v_{\mathrm{idx}_i}$ at those indices with a candidate item randomly selected from the corresponding candidate set $c_{\mathrm{idx}_i}$. The above augmentation process can be formulated as:
\begin{equation}
s_u^{\prime}=\operatorname{T-Substitute}\left(s_u\right)=\left[v_1, v_2, \ldots, \bar{v}_{\mathrm{idx}_i}, \ldots, v_{\left|s_u\right|}\right],
\label{eq:T-Substitute}
\end{equation}
where $\bar{v}_{\mathrm{idx}_i} \in c_{\mathrm{idx}_i}$ is the substituted item. Unlike traditional operators that directly use $s_u^{\prime}$ as a new sample to participate in model training, we mixup the output sequence representation of $s_u$ and $s_u^{\prime}$ from the sequence encoder to generate new training samples in the representation space \cite{zhang2017mixup,verma2019manifold}:
\begin{equation}
E_u^{\prime} = \operatorname{Look-up}\left(\mathbf{M}_\mathcal{V}, s_u^{\prime}\right), \quad H_u^{\prime} = \operatorname{Encoder}\left(E_u^{\prime}\right),
\label{eq:mixing-a}
\end{equation}
\begin{equation}
H_u^{a} = \lambda \cdot H_u + (1-\lambda) \cdot H_u^{\prime},
\label{eq:mixing-b}
\end{equation}
where $\lambda \sim \operatorname{Beta}(\alpha, \alpha)$ is the mixup weight and $H_u^{ao}$ is final augmented representation used for model training. It has been demonstrated that hybrid operations can enhance sample novelty while preserving important semantic information \cite{dang2025augmenting}.

\vspace{0.3em}

\noindent \textbf{T-Insert.} Given an original sequence $s_u$, T-Insert iterates through all items in the sequence and selects each tail item with probability $p$. The indices of these selected items will be recorded as  $\left\{\operatorname{idx}_1, \operatorname{idx}_2, \ldots, \operatorname{idx}_i\right\}$. It adopts the same method as T-Substitute for drawing the $p$. Then, we insert an item before each selected tail item $v_{\mathrm{idx}_i}$ at those indices. The inserted item is randomly selected from the corresponding candidate set $c_{\mathrm{idx}_i}$. The above augmentation process can be formulated as:
\begin{equation}
s_u^{\prime}=\operatorname{T-Insert}\left(s_u\right)=\left[v_1, v_2, \ldots, \bar{v}_c,v_{\mathrm{idx}_i}, \ldots, v_{\left|s_u\right|}\right],
\label{eq:T-Insert-a}
\end{equation}
where $\bar{v}_c \in c_{\mathrm{idx}_i}$ is the inserted item. After insertion, the length of $s_u^{\prime}$ will exceed that of the original sequence $s_u$. Therefore, to enable proper mixing of $s_u^{\prime}$ and $s_u$, we extend the original sequence using the selected trailing items: 
\begin{equation}
s_u^{e}=\operatorname{Extend}\left(s_u\right)=\left[v_1, v_2, \ldots, v_{\mathrm{idx}_i},v_{\mathrm{idx}_i}, \ldots, v_{\left|s_u\right|}\right].
\label{eq:T-Insert-b}
\end{equation}
In other words, the item at the end of each insertion position will be duplicated once to ensure $|s_u^{\prime}| = |s_u^{e}|$. We follow the same procedure as for T-Substitute to mixup $s_u^{\prime}$ and $s_u^{e}$, yielding the final augmented representation $H_u^{ao}$. In  T-Insert, we insert relevant items before tail items to enable the model to learn the correlation between head and tail items more effectively when predicting the next item. Extending the sequence further increases the frequency of tail items appearing. Both approaches boost the interaction frequency for tail users.

\vspace{0.3em}

\noindent \textbf{Length-Based Operator Selection.} During the augmentation process, short tail user sequences should undergo more insertion operations to increase their interaction frequency further \cite{liu2021contrastive}. Therefore, we assign augmentation operators to each sequence with varying probabilities based on sequence length. Specifically, the probability of each sequence undergoing insertion augmentation is:
\begin{equation}
p_{insert} = 1 -(|s_u|/N), \quad p_{substitute} = 1 - p_{insert}
\label{eq:select}
\end{equation}
where $N$ is the max sequence length. Thus, short sequences will have a higher probability of undergoing insertion operations, while long sequences are more likely to undergo replacement operations.

\subsection{Tail-Aware Cross Augmentation}
Building upon the augmentation operator for single sequences, we aim to enable the model to learn further the associative knowledge and collaborative signals between different tail users. Therefore, we further extended the mixing operation to different sequences. Different from the previous unselected random mixing \cite{bian2022relevant,dang2025augmenting}, we distinguished between user sequences that prefer head items and those that prefer tail items. For each user sequence $s_u$, if the proportion of tail items exceeds $\beta$, we classify it as a sequence indicating a preference for tail items. Conversely, sequences where the proportion of tail items is below $\beta$ are classified as indicating a preference for head items. The reason for performing sequence classification here is that tail items and tail users do not necessarily appear simultaneously \cite{tan2025taming}. In other words, head users may purchase numerous tail items, while tail users may predominantly buy head items during their infrequent purchases. We aim to identify users genuinely interested in long-tail items for specialized tail-aware cross augmentation, thereby further improving long-tail performance.

Given a batch of sequences $\left\{s_u\right\}_{u=1}^B$, we can obtain the corresponding batch of representations $\left\{H_u\right\}_{u=1}^B$. The $B$ is the batch size. We first classify these sequences based on the proportion of tail items and obtain $\left\{H_u\right\}_{u=1}^{B_h}$ and $\left\{H_u\right\}_{u=1}^{B_t}$, where $B_h$ and $B_t$ are the number of head preferred sequences and tail preferred sequences. We use $\left\{H_u\right\}_{u=1}^{B_t}$ as an example to introduce cross augmentation. The $\left\{H_u\right\}_{u=1}^{B_h}$ will be augmented similarly.

We start by shuffling $\left\{H_u\right\}_{u=1}^{B_t}$ from the batch perspective \cite{verma2019manifold}:
\begin{equation}
\left\{H_u^{\prime}\right\}_{u=1}^{B_t} = \operatorname{Shuffle} \left\{H_u\right\}_{u=1}^{B_t}.
\label{eq:shuffle}
\end{equation}
After that, we sample mixup weights $\lambda_t = \left\{\lambda_1,\lambda_2,\dots,\lambda_{B_t}\right\}$ from the $\operatorname{Beta}(\alpha, \alpha)$.
The mixup is performed as follows:
\begin{equation}
\left\{H_u^{ac} \right\}_{u=1}^{B_t} = \lambda_t \cdot \left\{H_u\right\}_{u=1}^{B_t} + (1-\Lambda_I) \cdot \left\{H_u^{\prime}\right\}_{u=1}^{B_t}.
\label{eq:cross-mixing}
\end{equation}
The cross augmentation will simultaneously perform the same mixup process for the representations of positive and negative items \cite{dang2025augmenting}. The final output is used to calculate the recommendation loss. To further enhance the novelty of the augmented samples, we also incorporated $s_u^{\prime}$ from T-Substitute and T-Insert into cross augmentation. They will follow the same classification as their corresponding original sequences.

\subsection{Model Training}
We adopt the commonly used binary cross-entropy (BCE) loss to train the sequential recommendation models \cite{kang2018self,li2020time,yue2024linear}:
\begin{equation}
\begin{aligned}
\mathcal{L}_{main} & = \operatorname{BCE}\left(H_u, E_u^{+}, E_u^{-}\right) \\ & = 
- \left[\log \left(\sigma\left(H_u \cdot E_u^{+}\right)\right)+\log \left(1 -\sigma\left(H_u \cdot E_u^{-}\right)\right)\right],
\end{aligned}
\label{eq:main}
\end{equation}
where $H_u, E_u^{+}$ and $E_u^{-}$ denote the representations of the user, the positive and negative items, respectively. $\sigma()$ is the sigmoid function. For each sequence, we select an operator according to Equation (\ref{eq:select}) and obtain $H_u^{ao}$. The loss is calculated similar to above:
\begin{equation}
\mathcal{L}_{operator} = \operatorname{BCE} \left(H_u^{ao}, E_u^{+}, E_u^{-}\right).
\end{equation}
For cross augmentation, we use all mixed representations to calculate the objective function:
\begin{equation}
\mathcal{L}_{cross} = \operatorname{BCE} \left(H_u^{ac}, E_u^{ac+}, E_u^{ac-}\right).
\end{equation}
Following the previous work \cite{qiu2021u,liu2022multi}, we adopt a two-stage training strategy to avoid the noise augmented samples produced by mixing unoptimized embeddings. In the first stage, we follow the standard SR model training process, with the primary objective of facilitating the learning of high-quality item embeddings from original data. We use Equation (\ref{eq:main}) as the sole objective function. In the second stage, we employ the two augmentation operators and cross augmentation:
\begin{equation}
\mathcal{L} = \mathcal{L}_{main} + \mathcal{L}_{operator} + \mathcal{L}_{cross}.
\label{eq:main+aug}
\end{equation}
We do not set additional loss weights for $\mathcal{L}_{operator}$ and $\mathcal{L}_{cross}$ since we treat all augmented samples equally with the original samples. Adaptive weights for different samples are reserved for future work.

\begin{table*}[!t]
  \centering
  \caption{Performance comparison. The best performance is bolded and the second-best is underlined.}
  \vspace{-1em}
      \renewcommand\arraystretch{0.85}
        \setlength{\tabcolsep}{1.5mm}{
  \scalebox{0.70}{
    \begin{tabular}{cc|cccc|cc|cccc|cc|cccc}
    \toprule
    \multirow{2}[1]{*}{Dataset} & \multirow{2}[1]{*}{Method} & \multicolumn{4}{c|}{Overall}  & \multicolumn{2}{c|}{Head Item} & \multicolumn{4}{c|}{Tail Item} & \multicolumn{2}{c|}{Head User} & \multicolumn{4}{c}{Tail User} \\
          &       & H@10  & N@10  & H@20  & N@20  & H@10  & N@10  & H@10  & N@10  & H@20  & N@20  & H@10  & N@10  & H@10  & N@10  & H@20  & N@20 \\
    \midrule
    \multirow{24}[5]{*}{Toys} & GRU4Rec & 0.0364  & 0.0192  & 0.0580  & 0.0246  & 0.0581  & 0.0311  & 0.0230  & 0.0119  & 0.0355  & 0.0150  & \underline{0.0456}  & 0.0210  & 0.0341  & 0.0188  & 0.0531  & 0.0235  \\
          & CMR   & 0.0415  & 0.0214  & 0.0662  & 0.0276  & 0.0633  & 0.0320  & 0.0281  & 0.0150  & 0.0419  & 0.0184  & 0.0428  & 0.0208  & 0.0412  & 0.0216  & 0.0651  & 0.0276  \\
          & CMRSI & 0.0405  & 0.0202  & 0.0647  & 0.0286  & 0.0669  & 0.0345  & 0.0273  & 0.0156  & \underline{0.0425}  & 0.0194  & 0.0411  & 0.0197  & 0.0399  & 0.0207  & 0.0630  & 0.0269  \\
          & RepPad & \underline{0.0442}  & \underline{0.0243}  & \underline{0.0641}  & \underline{0.0293}  & 0.0692  & 0.0368  & \underline{0.0288}  & \underline{0.0166}  & 0.0406  & \underline{0.0196}  & 0.0368  & 0.0182  & \underline{0.0460}  & \underline{0.0258}  & \underline{0.0657}  & \underline{0.0308}  \\
          & CITIES & 0.0393  & 0.0211  & 0.0617  & 0.0268  & 0.0636  & 0.0339  & 0.0244  & 0.0132  & 0.0380  & 0.0167  & 0.0404  & \underline{0.0211}  & 0.0390  & 0.0211  & 0.0590  & 0.0262  \\
          & LOAM  & 0.0223  & 0.0116  & 0.0363  & 0.0151  & 0.0158  & 0.0076  & 0.0263  & 0.0141  & 0.0389  & 0.0172  & 0.0350  & 0.0178  & 0.0191  & 0.0101  & 0.0299  & 0.0128  \\
          & MELT  & 0.0330  & 0.0174  & 0.0525  & 0.0223  & \textbf{0.0819} & \textbf{0.0432} & 0.0029  & 0.0016  & 0.0046  & 0.0020  & 0.0265  & 0.0127  & 0.0346  & 0.0186  & 0.0539  & 0.0234  \\
          \rowcolor{mygray}\cellcolor{white} & \textbf{TADA} & \textbf{0.0512} & \textbf{0.0280} & \textbf{0.0749} & \textbf{0.0340} & \underline{0.0746}  & \underline{0.0412}  & \textbf{0.0368} & \textbf{0.0200} & \textbf{0.0513} & \textbf{0.0236} & \textbf{0.0515} & \textbf{0.0265} & \textbf{0.0511} & \textbf{0.0284} & \textbf{0.0722} & \textbf{0.0337} \\
\cmidrule{2-18}          & SASRec & 0.0716  & 0.0396  & 0.1015  & 0.0471  & 0.0892  & 0.0485  & 0.0607  & 0.0341  & 0.0821  & 0.0395  & 0.0675  & 0.0345  & 0.0726  & 0.0408  & 0.1009  & 0.0480  \\
          & CMR   & 0.0652  & 0.0361  & 0.0959  & 0.0438  & 0.0994  & 0.0557  & 0.0414  & 0.0241  & 0.0639  & 0.0290  & 0.0608  & 0.0350  & 0.0663  & 0.0364  & 0.0985  & 0.0438  \\
          & CMRSI & 0.0683  & 0.0380  & 0.0984  & 0.0452  & 0.1013  & 0.0579  & 0.0455  & 0.0260  & 0.0694  & 0.0308  & 0.0657  & 0.0374  & 0.0716  & 0.0385  & 0.1024  & 0.0459  \\
          & RepPad & \underline{0.0805}  & \underline{0.0476}  & \underline{0.1089}  & \underline{0.0547}  & \underline{0.1030}  & \textbf{0.0599} & \underline{0.0667}  & \underline{0.0399}  & \underline{0.0855}  & \underline{0.0447}  & \underline{0.0726}  & \underline{0.0408}  & \underline{0.0825}  & \underline{0.0479}  & \underline{0.1064}  & \underline{0.0548}  \\
          & CITIES & 0.0723  & 0.0392  & 0.1024  & 0.0468  & 0.0971  & 0.0525  & 0.0570  & 0.0310  & 0.0805  & 0.0370  & 0.0711  & 0.0353  & 0.0726  & 0.0402  & 0.1015  & 0.0475  \\
          & LOAM  & 0.0569  & 0.0321  & 0.0815  & 0.0383  & 0.0458  & 0.0225  & 0.0637  & 0.0380  & 0.0827  & 0.0428  & 0.0626  & 0.0333  & 0.0554  & 0.0318  & 0.0774  & 0.0374  \\
          & MELT  & 0.0432  & 0.0216  & 0.0707  & 0.0286  & 0.0940  & 0.0474  & 0.0119  & 0.0058  & 0.0218  & 0.0082  & 0.0309  & 0.0133  & 0.0462  & 0.0237  & 0.0742  & 0.0308  \\
          \rowcolor{mygray}\cellcolor{white} & \textbf{TADA} & \textbf{0.0846} & \textbf{0.0491} & \textbf{0.1142} & \textbf{0.0566} & \textbf{0.1059} & \underline{0.0597}  & \textbf{0.0716} & \textbf{0.0427} & \textbf{0.0927} & \textbf{0.0480} & \textbf{0.0824} & \textbf{0.0459} & \textbf{0.0852} & \textbf{0.0500} & \textbf{0.1115} & \textbf{0.0565} \\
\cmidrule{2-18}          & FMLP-Rec & 0.0555  & 0.0325  & 0.0745  & 0.0373  & 0.0719  & 0.0400  & 0.0454  & 0.0279  & 0.0553  & 0.0304  & 0.0546  & 0.0308  & 0.0557  & 0.0330  & 0.0728  & 0.0373  \\
          & CMR   & 0.0601  & 0.0342  & 0.0770  & 0.0380  & 0.0792  & 0.0441  & 0.0459  & 0.0283  & 0.0572  & 0.0306  & 0.0587  & 0.0326  & 0.0555  & 0.0339  & 0.0763  & 0.0366  \\
          & CMRSI & \underline{0.0623}  & \underline{0.0360}  & 0.0814  & \underline{0.0405}  & 0.0825  & 0.0463  & \underline{0.0489}  & 0.0295  & 0.0590  & 0.0322  & \underline{0.0603}  & \underline{0.0347}  & 0.0587  & 0.0355  & 0.0796  & 0.0387  \\
          & RepPad & 0.0562  & 0.0352  & 0.0778  & 0.0385  & 0.0861  & \underline{0.0515}  & 0.0410  & 0.0242  & 0.0542  & 0.0275  & 0.0556  & 0.0311  & 0.0575  & \underline{0.0362}  & 0.0774  & 0.0389  \\
          & CITIES & 0.0586  & 0.0334  & \underline{0.0832}  & 0.0396  & 0.0883  & 0.0511  & 0.0403  & 0.0225  & 0.0560  & 0.0265  & 0.0546  & 0.0279  & \underline{0.0596}  & 0.0347  & \underline{0.0840}  & \underline{0.0409}  \\
          & LOAM  & 0.0468  & 0.0266  & 0.0651  & 0.0312  & 0.0460  & 0.0211  & 0.0473  & \underline{0.0299}  & \underline{0.0596}  & \underline{0.0330}  & 0.0513  & 0.0273  & 0.0457  & 0.0264  & 0.0621  & 0.0305  \\
          & MELT  & 0.0351  & 0.0185  & 0.0549  & 0.0235  & \underline{0.0890}  & 0.0475  & 0.0020  & 0.0007  & 0.0048  & 0.0014  & 0.0309  & 0.0155  & 0.0362  & 0.0192  & 0.0563  & 0.0243  \\
          \rowcolor{mygray}\cellcolor{white} & \textbf{TADA} & \textbf{0.0664} & \textbf{0.0399} & \textbf{0.0869} & \textbf{0.0451} & \textbf{0.0883} & \textbf{0.0502} & \textbf{0.0528} & \textbf{0.0335} & \textbf{0.0653} & \textbf{0.0367} & \textbf{0.0649} & \textbf{0.0379} & \textbf{0.0667} & \textbf{0.0404} & \textbf{0.0853} & \textbf{0.0451} \\
    \midrule
    \multirow{24}[6]{*}{Beauty} & GRU4Rec & 0.0383  & 0.0194  & 0.0639  & 0.0258  & 0.0730  & 0.0364  & 0.0119  & 0.0064  & 0.0204  & 0.0086  & 0.0604  & 0.0306  & 0.0328  & 0.0165  & 0.0544  & 0.0220  \\
          & CMR   & 0.0449  & 0.0230  & 0.0750  & 0.0305  & 0.0813  & 0.0408  & 0.0173  & 0.0086  & 0.0263  & 0.0109  & 0.0711  & 0.0374  & 0.0384  & 0.0193  & 0.0641  & 0.0258  \\
          & CMRSI & \underline{0.0503}  & \underline{0.0254}  & \underline{0.0797}  & \underline{0.0329}  & \underline{0.0932}  & \underline{0.0469}  & 0.0176  & 0.0091  & 0.0278  & 0.0111  & \underline{0.0756}  & \underline{0.0382}  & \underline{0.0437}  & \underline{0.0220}  & \underline{0.0702}  & \underline{0.0287}  \\
          & RepPad & 0.0466  & 0.0243  & 0.0729  & 0.0309  & 0.0861  & 0.0449  & 0.0165  & 0.0083  & 0.0266  & 0.0110  & 0.0635  & 0.0318  & 0.0424  & 0.0214  & 0.0665  & 0.0281  \\
          & CITIES & 0.0394  & 0.0199  & 0.0655  & 0.0264  & 0.0741  & 0.0375  & 0.0130  & 0.0064  & 0.0220  & 0.0086  & 0.0613  & 0.0316  & 0.0340  & 0.0169  & 0.0561  & 0.0225  \\
          & LOAM  & 0.0283  & 0.0135  & 0.0465  & 0.0180  & 0.0411  & 0.0188  & \underline{0.0184}  & \underline{0.0094}  & \underline{0.0293}  & \underline{0.0121}  & 0.0543  & 0.0253  & 0.0217  & 0.0105  & 0.0369  & 0.0143  \\
          & MELT  & 0.0398  & 0.0191  & 0.0670  & 0.0259  & 0.0912  & 0.0434  & 0.0006  & 0.0005  & 0.0019  & 0.0008  & 0.0470  & 0.0220  & 0.0380  & 0.0184  & 0.0636  & 0.0247  \\
          \rowcolor{mygray}\cellcolor{white} & \textbf{TADA} & \textbf{0.0555} & \textbf{0.0279} & \textbf{0.0836} & \textbf{0.0350} & \textbf{0.1027} & \textbf{0.0515} & \textbf{0.0195} & \textbf{0.0100} & \textbf{0.0321} & \textbf{0.0131} & \textbf{0.0823} & \textbf{0.0409} & \textbf{0.0488} & \textbf{0.0247} & \textbf{0.0720} & \textbf{0.0305} \\
\cmidrule{2-18}          & SASRec & 0.0605  & 0.0318  & 0.0918  & 0.0397  & 0.0999  & 0.0507  & 0.0304  & 0.0174  & 0.0452  & 0.0211  & 0.0825  & 0.0420  & 0.0549  & 0.0292  & 0.0842  & 0.0366  \\
          & CMR   & 0.0580  & 0.0306  & 0.0916  & 0.0391  & 0.1067  & 0.0559  & 0.0210  & 0.0114  & 0.0343  & 0.0147  & 0.0774  & 0.0396  & 0.0532  & 0.0284  & 0.0856  & 0.0367  \\
          & CMRSI & 0.0594  & 0.0315  & 0.0936  & 0.0401  & 0.1079  & 0.0561  & 0.0224  & 0.0127  & 0.0358  & 0.0161  & 0.0704  & 0.0383  & 0.0566  & 0.0298  & 0.0888  & 0.0379  \\
          & RepPad & \underline{0.0715}  & \underline{0.0391}  & \underline{0.1002}  & \underline{0.0479}  & \textbf{0.1205} & \textbf{0.0652} & 0.0363  & 0.0210  & \underline{0.0542}  & 0.0258  & \textbf{0.1004} & \textbf{0.0541} & \underline{0.0648}  & \underline{0.0355}  & \underline{0.0958}  & \underline{0.0432}  \\
          & CITIES & 0.0652  & 0.0342  & 0.0941  & 0.0415  & 0.1059  & 0.0554  & 0.0342  & 0.0181  & 0.0479  & 0.0215  & 0.0946  & 0.0517  & 0.0579  & 0.0299  & 0.0836  & 0.0364  \\
          & LOAM  & 0.0489  & 0.0266  & 0.0763  & 0.0335  & 0.0633  & 0.0310  & \underline{0.0380}  & \underline{0.0232}  & 0.0534  & \underline{0.0271}  & 0.0783  & 0.0434  & 0.0416  & 0.0224  & 0.0660  & 0.0285  \\
          & MELT  & 0.0469  & 0.0224  & 0.0772  & 0.0300  & 0.1078  & 0.0516  & 0.0005  & 0.0002  & 0.0020  & 0.0006  & 0.0570  & 0.0278  & 0.0444  & 0.0211  & 0.0733  & 0.0283  \\
          \rowcolor{mygray}\cellcolor{white} & \textbf{TADA} & \textbf{0.0732} & \textbf{0.0407} & \textbf{0.1076} & \textbf{0.0493} & \underline{0.1157}  & \underline{0.0612}  & \textbf{0.0409} & \textbf{0.0251} & \textbf{0.0593} & \textbf{0.0297} & \underline{0.0946}  & \underline{0.0517}  & \textbf{0.0679} & \textbf{0.0379} & \textbf{0.0982} & \textbf{0.0455} \\
\cmidrule{2-18}          & FMLP-Rec & 0.0539  & 0.0289  & 0.0774  & 0.0349  & 0.0891  & 0.0465  & 0.0270  & 0.0155  & 0.0338  & 0.0173  & 0.0803  & 0.0440  & 0.0473  & 0.0251  & 0.0677  & 0.0303  \\
          & CMR   & 0.0594  & 0.0324  & 0.0803  & 0.0386  & \underline{0.1061}  & 0.0552  & 0.0211  & 0.0111  & 0.0363  & 0.0149  & 0.0758  & 0.0407  & 0.0534  & 0.0199  & 0.0628  & 0.0227  \\
          & CMRSI & \underline{0.0612}  & 0.0335  & 0.0825  & \underline{0.0409}  & 0.1023  & 0.0586  & 0.0219  & 0.0115  & 0.0354  & 0.0147  & 0.0770  & 0.0405  & 0.0542  & 0.0209  & 0.0640  & 0.0234  \\
          & RepPad & 0.0604  & \underline{0.0344}  & \underline{0.0838}  & 0.0403  & 0.1053  & \underline{0.0601}  & 0.0262  & 0.0148  & 0.0356  & 0.0172  & \underline{0.0832}  & \underline{0.0498}  & \underline{0.0547}  & \underline{0.0306}  & \underline{0.0745}  & \underline{0.0357}  \\
          & CITIES & 0.0556  & 0.0301  & 0.0819  & 0.0367  & 0.1024  & 0.0546  & 0.0200  & 0.0115  & 0.0263  & 0.0130  & 0.0747  & 0.0422  & 0.0509  & 0.0271  & 0.0752  & 0.0332  \\
          & LOAM  & 0.0396  & 0.0213  & 0.0616  & 0.0268  & 0.0502  & 0.0238  & \underline{0.0314}  & \underline{0.0194}  & \underline{0.0409}  & \underline{0.0217}  & 0.0709  & 0.0378  & 0.0317  & 0.0172  & 0.0503  & 0.0218  \\
          & MELT  & 0.0475  & 0.0258  & 0.0728  & 0.0322  & 0.1061  & 0.0582  & 0.0028  & 0.0012  & 0.0056  & 0.0019  & 0.0586  & 0.0334  & 0.0447  & 0.0239  & 0.0678  & 0.0297  \\
          \rowcolor{mygray}\cellcolor{white} & \textbf{TADA} & \textbf{0.0658} & \textbf{0.0384} & \textbf{0.0917} & \textbf{0.0450} & \textbf{0.1099} & \textbf{0.0609} & \textbf{0.0322} & \textbf{0.0213} & \textbf{0.0429} & \textbf{0.0240} & \textbf{0.0977} & \textbf{0.0563} & \textbf{0.0579} & \textbf{0.0340} & \textbf{0.0812} & \textbf{0.0399} \\
    \midrule
    \multirow{24}[6]{*}{Sports} & GRU4Rec & 0.0155  & 0.0073  & 0.0285  & 0.0105  & 0.0289  & 0.0135  & 0.0052  & 0.0025  & 0.0092  & 0.0035  & 0.0173  & 0.0081  & 0.0150  & 0.0071  & 0.0279  & 0.0103  \\
          & CMR   & 0.0220  & 0.0109  & 0.0383  & 0.0150  & 0.0410  & 0.0194  & \underline{0.0080}  & \underline{0.0046}  & \underline{0.0121}  & \underline{0.0054}  & 0.0233  & 0.0109  & 0.0221  & 0.0111  & 0.0386  & 0.0152  \\
          & CMRSI & 0.0240  & 0.0119  & 0.0414  & 0.0159  & 0.0468  & 0.0233  & 0.0069  & 0.0035  & 0.0115  & 0.0047  & 0.0216  & 0.0105  & \underline{0.0250}  & \underline{0.0125}  & 0.0412  & \underline{0.0164}  \\
          & RepPad & 0.0224  & 0.0111  & 0.0364  & 0.0136  & 0.0431  & 0.0212  & 0.0065  & 0.0033  & 0.0109  & 0.0044  & 0.0212  & 0.0103  & 0.0227  & 0.0105  & 0.0368  & 0.0148  \\
          & CITIES & 0.0179  & 0.0084  & 0.0313  & 0.0118  & 0.0339  & 0.0159  & 0.0056  & 0.0027  & 0.0089  & 0.0035  & 0.0197  & 0.0087  & 0.0175  & 0.0083  & 0.0310  & 0.0117  \\
          & LOAM  & 0.0155  & 0.0079  & 0.0258  & 0.0105  & 0.0280  & 0.0144  & 0.0059  & 0.0029  & 0.0097  & 0.0039  & 0.0155  & 0.0074  & 0.0155  & 0.0080  & 0.0256  & 0.0105  \\
          & MELT  & \underline{0.0246}  & \underline{0.0120}  & \underline{0.0421}  & \underline{0.0164}  & \textbf{0.0564}  & \underline{0.0276}  & 0.0001  & 0.0000  & 0.0001  & 0.0000  & \underline{0.0256}  & \underline{0.0133}  & 0.0243  & 0.0117  & \underline{0.0418}  & 0.0161  \\
          \rowcolor{mygray}\cellcolor{white} & \textbf{TADA} & \textbf{0.0279} & \textbf{0.0140} & \textbf{0.0450} & \textbf{0.0183} & \underline{0.0519} & \textbf{0.0258} & \textbf{0.0095} & \textbf{0.0049} & \textbf{0.0159} & \textbf{0.0065} & \textbf{0.0306} & \textbf{0.0148} & \textbf{0.0273} & \textbf{0.0138} & \textbf{0.0438} & \textbf{0.0179} \\
\cmidrule{2-18}          & SASRec & 0.0317  & 0.0177  & 0.0479  & 0.0218  & 0.0488  & 0.0260  & 0.0185  & 0.0113  & 0.0248  & 0.0129  & 0.0301  & 0.0160  & 0.0321  & 0.0181  & 0.0485  & 0.0222  \\
          & CMR   & 0.0336  & 0.0182  & 0.0550  & 0.0236  & 0.0673  & 0.0367  & 0.0076  & 0.0040  & 0.0151  & 0.0059  & 0.0315  & 0.0175  & 0.0332  & 0.0184  & 0.0554  & 0.0238  \\
          & CMRSI & 0.0326  & 0.0180  & 0.0493  & 0.0222  & 0.0655  & 0.0360  & 0.0072  & 0.0042  & 0.0115  & 0.0052  & 0.0310  & 0.0173  & 0.0322  & 0.0172  & 0.0504  & 0.0226  \\
          & RepPad & \underline{0.0379}  & \underline{0.0205}  & \underline{0.0589} & \underline{0.0248} & \underline{0.0732}  & \underline{0.0362}  & 0.0182  & 0.0099  & 0.0245  & 0.0119  & \underline{0.0370}  & \underline{0.0196}  & \underline{0.0379}  & \underline{0.0215}  & \underline{0.0583}  & \underline{0.0258}  \\
          & CITIES & 0.0334  & 0.0187  & 0.0504  & 0.0229  & 0.0555  & 0.0302  & 0.0164  & 0.0098  & 0.0228  & 0.0114  & 0.0309  & 0.0171  & 0.0340  & 0.0190  & 0.0511  & 0.0233  \\
          & LOAM  & 0.0212  & 0.0111  & 0.0333  & 0.0142  & 0.0222  & 0.0106  & \underline{0.0204}  & \underline{0.0116}  & \underline{0.0266}  & \underline{0.0131}  & 0.0192  & 0.0102  & 0.0217  & 0.0114  & 0.0338  & 0.0144  \\
          & MELT  & 0.0326  & 0.0168  & 0.0513  & 0.0214  & \textbf{0.0748} & \textbf{0.0385} & 0.0001  & 0.0000  & 0.0002  & 0.0001  & 0.0320  & 0.0171  & 0.0327  & 0.0167  & 0.0512  & 0.0213  \\
          \rowcolor{mygray}\cellcolor{white} & \textbf{TADA} & \textbf{0.0402} & \textbf{0.0217} & \textbf{0.0605} & \textbf{0.0267} & 0.0654  & 0.0338  & \textbf{0.0208} & \textbf{0.0123} & \textbf{0.0299} & \textbf{0.0146} & \textbf{0.0382} & \textbf{0.0209} & \textbf{0.0407} & \textbf{0.0219} & \textbf{0.0611} & \textbf{0.0270} \\
\cmidrule{2-18}          & FMLP-Rec & 0.0242  & 0.0122  & 0.0365  & 0.0153  & 0.0404  & 0.0186  & 0.0118  & 0.0073  & 0.0149  & 0.0081  & 0.0202  & 0.0095  & 0.0252  & 0.0129  & 0.0372  & 0.0159  \\
          & CMR   & 0.0257  & 0.0129  & 0.0394  & 0.0163  & 0.0432  & 0.0199  & 0.0112  & 0.0065  & 0.0147  & 0.0078  & 0.0220  & 0.0102  & 0.0259  & 0.0140  & 0.0383  & 0.0164  \\
          & CMRSI & 0.0265  & 0.0138  & 0.0419  & 0.0178  & 0.0471  & 0.0224  & 0.0125  & 0.0078  & 0.0160  & 0.0092  & 0.0247  & 0.0118  & \underline{0.0279}  & 0.0148  & \underline{0.0419}  & 0.0178  \\
          & RepPad & 0.0272  & 0.0147  & 0.0415  & 0.0183  & 0.0522  & 0.0281  & 0.0105  & 0.0061  & 0.0147  & 0.0071  & 0.0277  & 0.0147  & 0.0269  & \underline{0.0152}  & 0.0401  & 0.0175  \\
          & CITIES & 0.0239  & 0.0123  & 0.0353  & 0.0152  & 0.0405  & 0.0208  & 0.0111  & 0.0058  & 0.0154  & 0.0069  & 0.0229  & 0.0114  & 0.0241  & 0.0125  & 0.0353  & 0.0154  \\
          & LOAM  & 0.0169  & 0.0092  & 0.0259  & 0.0115  & 0.0197  & 0.0094  & \underline{0.0147}  & \underline{0.0091}  & \underline{0.0183}  & \underline{0.0101}  & 0.0171  & 0.0088  & 0.0168  & 0.0094  & 0.0255  & 0.0115  \\
          & MELT  & \underline{0.0278}  & \underline{0.0150}  & \underline{0.0422}  & \underline{0.0186}  & \textbf{0.0632} & \textbf{0.0342} & 0.0005  & 0.0002  & 0.0012  & 0.0003  & \underline{0.0294}  & \underline{0.0164}  & 0.0274  & 0.0146  & 0.0414  & \underline{0.0181}  \\
          \rowcolor{mygray}\cellcolor{white} & \textbf{TADA} & \textbf{0.0304} & \textbf{0.0161} & \textbf{0.0450} & \textbf{0.0198} & \underline{0.0483}  & \underline{0.0240} & \textbf{0.0166} & \textbf{0.0100} & \textbf{0.0207} & \textbf{0.0111} & \textbf{0.0298} & \textbf{0.0151} & \textbf{0.0305} & \textbf{0.0164} & \textbf{0.0448} & \textbf{0.0200} \\
    \bottomrule
    \end{tabular}}}%
  \label{tab:main}%
  \vspace{-1em}
\end{table*}%

\subsection{Discussion}
Compared to existing long-tail SR methods and data augmentation methods, our approach has the following advantages:

\begin{itemize}
    \item \textbf{Mitigating tail data sparsity:} Existing long-tail methods primarily rely on knowledge transfer from the head, struggling to effectively increase interaction counts in the tail, which results in limited performance gains or seesaw issues. Our TADA approach tailored augmentation methods for tail users and items, effectively improving the model's learning of tail representations while maintaining overall and head performance.
    \item \textbf{Low computational complexity:} Since we pre-construct the candidate set by linear models before augmentation begins, there is no need to select items based on similarity during augmentation, which would introduce $O(|\mathcal{V}|^2)$ complexity \cite{liu2021contrastive}. During the augmentation process, item selection, weight sampling, and representation mixing all exhibit linear complexity.
    \item \textbf{High generalization ability:} Some augmentation methods leveraged bidirectional Transformers \cite{liu2021augmenting} or diffusion models \cite{liu2023diffusion}, which have limitations on the type of backbone network. Some long-tail methods also require the introduction of additional learnable modules \cite{kim2023melt}. In contrast, our method is model agnostic with no additional learnable parameters during training. 
\end{itemize}

\section{Experiments}

\subsection{Experimental Setup}

\begin{table*}[!t]
  \centering
  \caption{Results on more backbone sequential recommendation models.}
  \vspace{-1em}
      \renewcommand\arraystretch{0.85}
        \setlength{\tabcolsep}{1.5mm}{
  \scalebox{0.70}{
    \begin{tabular}{cc|cccc|cc|cccc|cc|cccc}
    \toprule
    \multirow{2}[1]{*}{Dataset} & \multirow{2}[1]{*}{Method} & \multicolumn{4}{c|}{Overall}  & \multicolumn{2}{c|}{Head Item} & \multicolumn{4}{c|}{Tail Item} & \multicolumn{2}{c|}{Head User} & \multicolumn{4}{c}{Tail User} \\
          &       & H@10  & N@10  & H@20  & N@20  & H@10  & N@10  & H@10  & N@10  & H@20  & N@20  & H@10  & N@10  & H@10  & N@10  & H@20  & N@20 \\
    \midrule
    \multirow{6}[0]{*}{Toys} & NARM  & 0.0371  & 0.0205  & 0.0562  & 0.0253  & 0.0583  & 0.0328  & 0.0241  & 0.0130  & 0.0350  & 0.0158  & 0.0160  & 0.0076  & 0.0424  & 0.0238  & 0.0625  & 0.0288  \\
          & \cellcolor{mygray}\textbf{w/ TADA} & \cellcolor{mygray}\textbf{0.0500} & \cellcolor{mygray}\textbf{0.0280} & \cellcolor{mygray}\textbf{0.0725} & \cellcolor{mygray}\textbf{0.0337} & \cellcolor{mygray}\textbf{0.0725} & \cellcolor{mygray}\textbf{0.0403} & \cellcolor{mygray}\textbf{0.0361} & \cellcolor{mygray}\textbf{0.0205} & \cellcolor{mygray}\textbf{0.0496} & \cellcolor{mygray}\textbf{0.0238} & \cellcolor{mygray}\textbf{0.0255} & \cellcolor{mygray}\textbf{0.0122} & \cellcolor{mygray}\textbf{0.0561} & \cellcolor{mygray}\textbf{0.0320} & \cellcolor{mygray}\textbf{0.0801} & \cellcolor{mygray}\textbf{0.0380} \\
          & LightSANs & 0.0501  & 0.0297  & 0.0688  & 0.0344  & 0.0632  & 0.0348  & 0.0421  & 0.0266  & 0.0517  & 0.0291  & 0.0482  & 0.0276  & 0.0506  & 0.0303  & 0.0685  & 0.0348  \\
          & \cellcolor{mygray}\textbf{w/ TADA} & \cellcolor{mygray}\textbf{0.0649} & \cellcolor{mygray}\textbf{0.0396} & \cellcolor{mygray}\textbf{0.0833} & \cellcolor{mygray}\textbf{0.0443} & \cellcolor{mygray}\textbf{0.0830} & \cellcolor{mygray}\textbf{0.0479} & \cellcolor{mygray}\textbf{0.0537} & \cellcolor{mygray}\textbf{0.0345} & \cellcolor{mygray}\textbf{0.0653} & \cellcolor{mygray}\textbf{0.0374} & \cellcolor{mygray}\textbf{0.0600} & \cellcolor{mygray}\textbf{0.0354} & \cellcolor{mygray}\textbf{0.0661} & \cellcolor{mygray}\textbf{0.0407} & \cellcolor{mygray}\textbf{0.0822} & \cellcolor{mygray}\textbf{0.0448} \\
          & gMLP  & 0.0482  & 0.0309  & 0.0619  & 0.0343  & 0.0607  & 0.0351  & 0.0404  & 0.0283  & 0.0483  & 0.0303  & 0.0541  & 0.0318  & 0.0467  & 0.0307  & 0.0593  & 0.0339  \\
          & \cellcolor{mygray}\textbf{w/ TADA} & \cellcolor{mygray}\textbf{0.0660} & \cellcolor{mygray}\textbf{0.0419} & \cellcolor{mygray}\textbf{0.0844} & \cellcolor{mygray}\textbf{0.0465} & \cellcolor{mygray}\textbf{0.0884} & \cellcolor{mygray}\textbf{0.0521} & \cellcolor{mygray}\textbf{0.0522} & \cellcolor{mygray}\textbf{0.0356} & \cellcolor{mygray}\textbf{0.0631} & \cellcolor{mygray}\textbf{0.0383} & \cellcolor{mygray}\textbf{0.0680} & \cellcolor{mygray}\textbf{0.0415} & \cellcolor{mygray}\textbf{0.0655} & \cellcolor{mygray}\textbf{0.0420} & \cellcolor{mygray}\textbf{0.0826} & \cellcolor{mygray}\textbf{0.0463} \\
    \midrule
    \multirow{6}[1]{*}{Beauty} & NARM  & 0.0410  & 0.0206  & 0.0649  & 0.0266  & 0.0803  & 0.0399  & 0.0111  & 0.0059  & 0.0170  & 0.0074  & 0.0403  & 0.0199  & 0.0412  & 0.0208  & 0.0639  & 0.0265  \\
          & \cellcolor{mygray}\textbf{w/ TADA} & \cellcolor{mygray}\textbf{0.0494} & \cellcolor{mygray}\textbf{0.0255} & \cellcolor{mygray}\textbf{0.0768} & \cellcolor{mygray}\textbf{0.0324} & \cellcolor{mygray}\textbf{0.0906} & \cellcolor{mygray}\textbf{0.0457} & \cellcolor{mygray}\textbf{0.0180} & \cellcolor{mygray}\textbf{0.0101} & \cellcolor{mygray}\textbf{0.0284} & \cellcolor{mygray}\textbf{0.0127} & \cellcolor{mygray}\textbf{0.0501} & \cellcolor{mygray}\textbf{0.0263} & \cellcolor{mygray}\textbf{0.0492} & \cellcolor{mygray}\textbf{0.0253} & \cellcolor{mygray}\textbf{0.0765} & \cellcolor{mygray}\textbf{0.0321} \\
          & LightSANs & 0.0566  & 0.0307  & 0.0788  & 0.0364  & 0.0964  & 0.0508  & 0.0262  & 0.0155  & 0.0345  & 0.0175  & 0.0872  & 0.0463  & 0.0489  & 0.0269  & 0.0686  & 0.0318  \\
          & \cellcolor{mygray}\textbf{w/ TADA} & \cellcolor{mygray}\textbf{0.0630} & \cellcolor{mygray}\textbf{0.0356} & \cellcolor{mygray}\textbf{0.0884} & \cellcolor{mygray}\textbf{0.0420} & \cellcolor{mygray}\textbf{0.1065} & \cellcolor{mygray}\textbf{0.0579} & \cellcolor{mygray}\textbf{0.0298} & \cellcolor{mygray}\textbf{0.0186} & \cellcolor{mygray}\textbf{0.0418} & \cellcolor{mygray}\textbf{0.0216} & \cellcolor{mygray}\textbf{0.0937} & \cellcolor{mygray}\textbf{0.0544} & \cellcolor{mygray}\textbf{0.0553} & \cellcolor{mygray}\textbf{0.0309} & \cellcolor{mygray}\textbf{0.0771} & \cellcolor{mygray}\textbf{0.0364} \\
          & gMLP  & 0.0473  & 0.0282  & 0.0657  & 0.0328  & 0.0811  & 0.0449  & 0.0216  & 0.0155  & 0.0268  & 0.0168  & 0.0725  & 0.0435  & 0.0410  & 0.0243  & 0.0562  & 0.0282  \\
          & \cellcolor{mygray}\textbf{w/ TADA} & \cellcolor{mygray}\textbf{0.0658} & \cellcolor{mygray}\textbf{0.0383} & \cellcolor{mygray}\textbf{0.0910} & \cellcolor{mygray}\textbf{0.0446} & \cellcolor{mygray}\textbf{0.1128} & \cellcolor{mygray}\textbf{0.0625} & \cellcolor{mygray}\textbf{0.0299} & \cellcolor{mygray}\textbf{0.0198} & \cellcolor{mygray}\textbf{0.0396} & \cellcolor{mygray}\textbf{0.0222} & \cellcolor{mygray}\textbf{0.0995} & \cellcolor{mygray}\textbf{0.0575} & \cellcolor{mygray}\textbf{0.0573} & \cellcolor{mygray}\textbf{0.0335} & \cellcolor{mygray}\textbf{0.0798} & \cellcolor{mygray}\textbf{0.0391} \\
    \midrule
    \multirow{6}[2]{*}{Sports} & NARM  & 0.0213  & 0.0105  & 0.0360  & 0.0142  & 0.0442  & 0.0220  & 0.0037  & 0.0017  & 0.0067  & 0.0025  & 0.0209  & 0.0098  & 0.0214  & 0.0107  & 0.0362  & 0.0144  \\
           & \cellcolor{mygray}\textbf{w/ TADA} & \cellcolor{mygray}\textbf{0.0297} & \cellcolor{mygray}\textbf{0.0155} & \cellcolor{mygray}\textbf{0.0455} & \cellcolor{mygray}\textbf{0.0195} & \cellcolor{mygray}\textbf{0.0584} & \cellcolor{mygray}\textbf{0.0306} & \cellcolor{mygray}\textbf{0.0077} & \cellcolor{mygray}\textbf{0.0039} & \cellcolor{mygray}\textbf{0.0110} & \cellcolor{mygray}\textbf{0.0048} & \cellcolor{mygray}\textbf{0.0284} & \cellcolor{mygray}\textbf{0.0157} & \cellcolor{mygray}\textbf{0.0301} & \cellcolor{mygray}\textbf{0.0155} & \cellcolor{mygray}\textbf{0.0463} & \cellcolor{mygray}\textbf{0.0196} \\
          & LightSANs & 0.0212  & 0.0116  & 0.0319  & 0.0143  & 0.0353  & 0.0184  & 0.0103  & 0.0064  & 0.0129  & 0.0070  & 0.0214  & 0.0112  & 0.0212  & 0.0117  & 0.0320  & 0.0144  \\
         & \cellcolor{mygray}\textbf{w/ TADA} & \cellcolor{mygray}\textbf{0.0293} & \cellcolor{mygray}\textbf{0.0160} & \cellcolor{mygray}\textbf{0.0421} & \cellcolor{mygray}\textbf{0.0192} & \cellcolor{mygray}\textbf{0.0499} & \cellcolor{mygray}\textbf{0.0256} & \cellcolor{mygray}\textbf{0.0134} & \cellcolor{mygray}\textbf{0.0086} & \cellcolor{mygray}\textbf{0.0175} & \cellcolor{mygray}\textbf{0.0097} & \cellcolor{mygray}\textbf{0.0295} & \cellcolor{mygray}\textbf{0.0149} & \cellcolor{mygray}\textbf{0.0292} & \cellcolor{mygray}\textbf{0.0163} & \cellcolor{mygray}\textbf{0.0416} & \cellcolor{mygray}\textbf{0.0194} \\
          & gMLP  & 0.0217  & 0.0119  & 0.0298  & 0.0139  & 0.0367  & 0.0187  & 0.0101  & 0.0066  & 0.0123  & 0.0072  & 0.0199  & 0.0107  & 0.0221  & 0.0122  & 0.0303  & 0.0142  \\
          & \cellcolor{mygray}\textbf{w/ TADA} & \cellcolor{mygray}\textbf{0.0317} & \cellcolor{mygray}\textbf{0.0181} & \cellcolor{mygray}\textbf{0.0454} & \cellcolor{mygray}\textbf{0.0215} & \cellcolor{mygray}\textbf{0.0567} & \cellcolor{mygray}\textbf{0.0310} & \cellcolor{mygray}\textbf{0.0125} & \cellcolor{mygray}\textbf{0.0082} & \cellcolor{mygray}\textbf{0.0167} & \cellcolor{mygray}\textbf{0.0092} & \cellcolor{mygray}\textbf{0.0324} & \cellcolor{mygray}\textbf{0.0177} & \cellcolor{mygray}\textbf{0.0315} & \cellcolor{mygray}\textbf{0.0182} & \cellcolor{mygray}\textbf{0.0452} & \cellcolor{mygray}\textbf{0.0216} \\
    \bottomrule
    \end{tabular}}}%
  \label{tab:more_backbones}%
  \vspace{-1em}
\end{table*}%

\noindent \textbf{Baselines.} \emph{(1) Backbone SR models:} To validate the generalizability of our proposed TADA, we first selected three representative backbone SR models: \textbf{GRU4Rec} \cite{hidasi2015session}, \textbf{SASRec} \cite{kang2018self}, and \textbf{FMLP-Rec} \cite{zhou2022filter}. These three models are based on recurrent neural networks (RNNs), attention mechanisms \cite{vaswani2017attention}, and multilayer perceptrons (MLPs), respectively. \emph{(2) Data augmentation methods:} \textbf{CMR} \cite{xie2022contrastive}, \textbf{CMRSI} \cite{liu2021contrastive}, and \textbf{RepPad} \cite{dang2024repeated}. We do not select augmentation methods based on contrastive learning \cite{xie2022contrastive,tian2023periodicity}, as these introduce additional learning objectives. In contrast, all augmentation data generated by our TADA is utilized for the main recommendation task. DiffASR \cite{liu2023diffusion} and ASReP \cite{liu2021augmenting} are not involved since they have limitations on the type of backbone models. Meanwhile, RepPad has reported better performance than them \cite{dang2024repeated}. Therefore, we chose RepPad as the baseline. \emph{(3) Long-tail SR methods}: \textbf{CITIES} \cite{jang2020cities}, \textbf{LOAM} \cite{yang2023loam}, and \textbf{MELT} \cite{kim2023melt}. To ensure a fair comparison, we do not include LLM-ESR \cite{liu2024llm} and R$^2$REC \cite{li2025reembedding}, as they use additional modal information, such as text and images, while our TADA only takes the item ID as input. Details about baselines are provided in the Appendix.

\vspace{0.3em}

\noindent \textbf{Datasets.} Following previous works \cite{li2025reembedding,dang2024repeated}, we select three datasets, \textbf{Toys}, \textbf{Beauty}, and \textbf{Sports} \cite{mcauley2015inferring}. We reduce the data by extracting the 5-core \cite{liu2021contrastive,dang2023uniform,tian2023periodicity}. The maximum sequence length is set to 50.

\vspace{0.3em}

\noindent \textbf{Evaluation Settings.} We adopt the leave-one-out strategy to partition sequences into training, validation, and test sets. Following previous practice \cite{xie2022contrastive,fan2022sequential,zhou2022filter,yue2024linear}, we rank the prediction over the whole item set rather than negative sampling, otherwise leading to biased discoveries \cite{krichene2020sampled}. The evaluation metrics include Hit Ratio@K (H@K) and Normalized Discounted Cumulative Gain@K (N@K). We report results with K $\in \{5, 10, 20\}$.

\vspace{0.3em}

\noindent \textbf{Implementation Details.} For all baselines, we adopt the implementation provided by the authors. We set the embedding size to 64 and the batch size to 256. To ensure fair comparisons, we carefully set and tune all other hyperparameters of each method as reported and suggested in the original papers. We use the Adam \cite{2014Adam} optimizer with the learning rate 0.001, $\beta_1=0.9$, $\beta_2=0.999$. For TADA, we tune the $\alpha, a, b, \beta$ in the range of $\{0.1,0.2,0.3,0.4,0.5\}$, $\{0.1,0.2,0.3\}$, $\{0.5,0.6,0.7,0.8\}$, $\{0.4,0.5,0.6\}$, respectively. The $K$ for correlation candidate set is tuned in the range of $\{10,20,30,40,50\}$. We conduct five runs and report the average results for all methods.

\begin{table*}[!t]
  \centering
  \caption{The ablation study results on the Sports dataset with GRU4Rec as the backbone SR model.}
    \vspace{-1em}
      \renewcommand\arraystretch{0.85}
        \setlength{\tabcolsep}{1.45mm}{
  \scalebox{0.70}{
    \begin{tabular}{c|cccc|cc|cccc|cc|cccc}
    \toprule
    \multirow{2}[2]{*}{Method} & \multicolumn{4}{c|}{Overall}  & \multicolumn{2}{c|}{Head Item} & \multicolumn{4}{c|}{Tail Item} & \multicolumn{2}{c|}{Head User} & \multicolumn{4}{c}{Tail User} \\
          & H@10  & N@10  & H@20  & \multicolumn{1}{c|}{N@20} & H@10  & \multicolumn{1}{c|}{N@10} & H@10  & N@10  & H@20  & \multicolumn{1}{c|}{N@20} & H@10  & \multicolumn{1}{c|}{N@10} & H@10  & N@10  & H@20  & N@20 \\
    \midrule
    TADA  & \textbf{0.0279} & \textbf{0.0140} & \textbf{0.0450} & \multicolumn{1}{c|}{\textbf{0.0183}} & \textbf{0.0519} & \multicolumn{1}{c|}{\textbf{0.0258}} & \textbf{0.0095}  & 0.0049 & \textbf{0.0159}  & \multicolumn{1}{c|}{\textbf{0.0065}} & \textbf{0.0306}  & \underline{0.0148} & \textbf{0.0273} & \textbf{0.0138} & \textbf{0.0438} & \textbf{0.0179} \\
    \midrule
    w/o Correlation Relationships & \underline{0.0273}  & \underline{0.0139}  & \underline{0.0429}  & \underline{0.0178}  & \underline{0.0505}  & \underline{0.0251}  & \underline{0.0094} & \textbf{0.0052}  & 0.0144  & \underline{0.0064}  & \underline{0.0303} & \textbf{0.0153} & \underline{0.0265}  & \underline{0.0135}  & \underline{0.0420}  & \underline{0.0174}  \\
    w/o Co-occurrence Relationships & 0.0248  & 0.0123  & 0.0414  & 0.0165  & 0.0448  & 0.0218  & \underline{0.0094} & 0.0050  & \underline{0.0151} & \underline{0.0064}  & 0.0268  & 0.0130  & 0.0243  & 0.0121  & 0.0401  & 0.0161  \\
    w/o Tail-Aware Operators & 0.0231  & 0.0118  & 0.0373  & 0.0153  & 0.0415  & 0.0204  & 0.0089  & \underline{0.0051}  & 0.0132  & 0.0062  & 0.0261  & 0.0125  & 0.0224  & 0.0116  & 0.0361  & 0.0150  \\
    w/o Cross Augmentation & 0.0213  & 0.0102  & 0.0358  & 0.0138  & 0.0403  & 0.0194  & 0.0067  & 0.0031  & 0.0100  & 0.0040  & 0.0206  & 0.0100  & 0.0215  & 0.0102  & 0.0359  & 0.0139  \\
    \midrule
    GRU4Rec & 0.0155  & 0.0073  & 0.0285  & 0.0105  & 0.0289  & 0.0135  & 0.0052  & 0.0025  & 0.0092  & 0.0035  & 0.0173  & 0.0081  & 0.0150  & 0.0071  & 0.0279  & 0.0103  \\
    \bottomrule
    \end{tabular}}}%
  \label{tab:ablation}%
  \vspace{-0.5em}
\end{table*}%

\subsection{Overall Comparison}
The experimental results of our TADA and different baseline methods with various backbone SR models are presented in Table \ref{tab:main} \footnote{For MELT \cite{kim2023melt}, although we reproduced it by strictly adhering to the code provided by the authors and the parameter settings \& adjustment ranges reported in the original paper, the performance consistently exhibits anomalies, with low performance in certain cases. We also observe this in other published papers \cite{liu2024llm,li2025reembedding}.}. We have the following observations and conclusions: 

\begin{itemize}
    \item Among the three backbone models, SASRec demonstrated the best overall performance, further validating the powerful capabilities of transformers in modeling sequences. Among the baseline methods based on data augmentation, CMRSI and RepPad achieved competitive results in some cases. The former generates augmented data through multiple heuristic operators, while the latter fills idle input spaces by leveraging the original sequence. However, these approaches lack augmentation methods specifically designed for tail data, potentially failing to improve tail performance effectively and even exacerbating head-tail imbalance. Consequently, data augmentation-based methods often underperform on tail items and tail users in many cases.
    \item For long-tail SR methods, we observe that they struggle to achieve advantages in overall and head performance. In some cases, their performance even falls short of the original model. This phenomenon further validates the seesaw effect we highlighted: these methods compromise overall and head performance while transferring knowledge to the tail, thereby degrading user experience. Although they achieve competitive results for some tail users and tail items, their overall performance still lags behind data augmentation-based approaches. This indicates that the sparsity of tail data severely constrains the model's learning capacity.
    \item For our proposed TADA, it achieves the best or the second-best performance in nearly all cases. Specifically, for long-tail items and long-tail users, it achieves optimal performance across all cases. TADA effectively mitigates data sparsity in the tail through its crafted tail-aware augmentation operator and cross augmentation, while also avoiding the seesaw effect caused by knowledge transfer. It significantly improves tail performance while simultaneously enhancing overall and head performance.
\end{itemize}

\subsection{Generalizing to More Backbones}\label{sec:more_backbones}
In order to verify the generalizability of TADA, we added our operators to more SR backbones: \textbf{NARM} \cite{li2017neural}, \textbf{LightSANs} \cite{fan2021lighter}, and \textbf{gMLP} \cite{liu2021pay}. Details about these models are provided in the Appendix. The results are presented in Table \ref{tab:more_backbones}. We can observe that our TADA still effectively improves the accuracy for the overall, head, and tail. Across three datasets, TADA achieved average improvements of 34.0\%, 23.7\%, and 44.9\%, respectively. Specifically, for long-tail recommendations, the average improvements are 34.5\%, 30.4\%, and 51.5\%. These results further demonstrate the effectiveness and generalization capabilities of our approach.

\begin{figure*}[!t]
  \centering
  \includegraphics[scale=0.465]{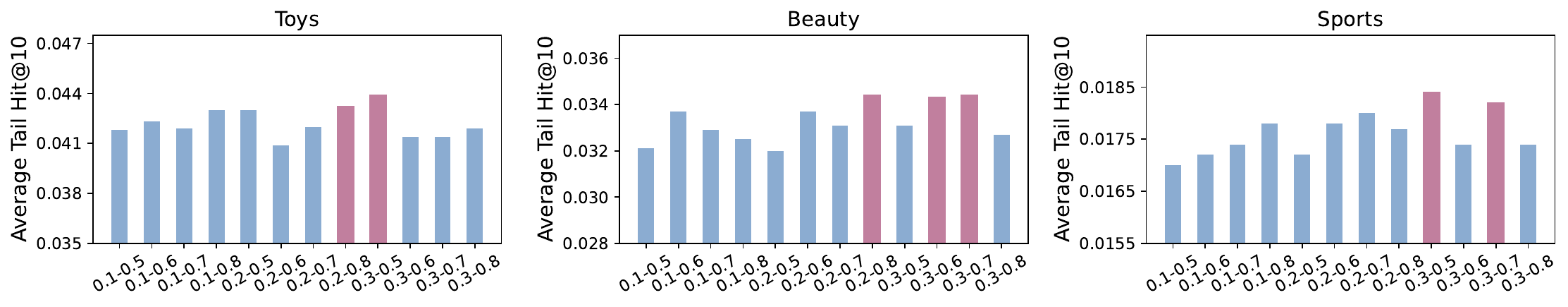}
  \vspace{-0.5em}
  \caption{Sensitivity analysis of hyperparameters $a$ and $b$. The `$0.1-0.5$' represents $a=0.1$ and $b=0.5$. We highlight the best and second-best results with different colors. The backbone is GRU4Rec. We report the average Hit@10 for tail items and users.}
  \label{fig:ab}
  \vspace{-0.5em}
\end{figure*}

\begin{figure*}[!t]
  \centering
  \includegraphics[scale=0.35]{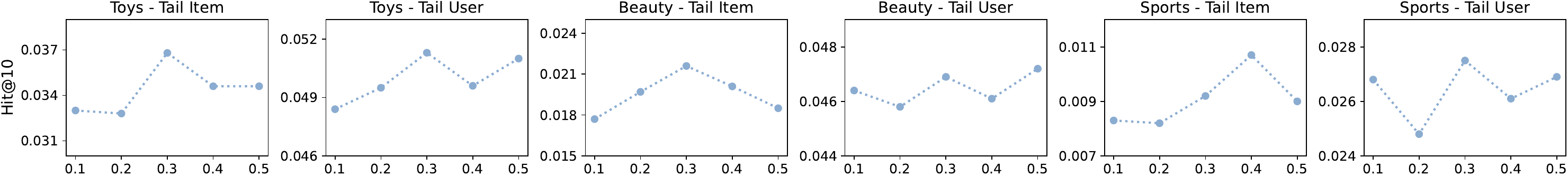}
  \vspace{-1em}
  \caption{Sensitivity analysis of hyperparameter $\alpha$ for Beta distribution.}
  \label{fig:beta}
  \vspace{-0.5em}
\end{figure*}

\subsection{Ablation Study}
We conduct an ablation study to verify the effectiveness of various components in our TADA. We compare the following variants: 1) w/o Correlation Relationships: Removing the correlation candidate set. 2) w/o Co-occurrence Relationships: Removing the co-occurrence candidate set. 3) w/o Tail-Aware Operators: Removing the proposed two tail-aware augmentation operators, T-Substitute and T-Insert. 4) w/o Cross Augmentation: Removing the tail-aware cross augmentation. The results are demonstrated in Table \ref{tab:ablation}.

We can observe that all modules contribute to performance improvements. For the correlation set and the co-occurrence set, they respectively contribute more to enhancing recommendation accuracy for tail items and tail users. The former provides augmentation candidates based on similarity, while the latter offers candidates based on co-occurrence relationships of items. After removing the two tail-aware augmentation operators and cross augmentation, the model's performance declined significantly. The augmentation operators contributed more to improving performance for tail users, while cross augmentation primarily enhanced the accuracy of tail items. Additionally, both significantly boosted overall performance and head performance, achieving a win-win scenario.

\subsection{Hyperparameter Investigation}
We further investigate the hyperparameters $a,b$ of the tail-aware operators and $\alpha$ of the Beta distribution. The results are presented in Figures \ref{fig:ab} and \ref{fig:beta}. For simplicity, in our TADA, the two operators share the same $a$ and $b$, and all the Beta distributions share the same $\alpha$. For $a$ and $b$, we observe optimal and suboptimal values occurring at $a=0.2, b=0.8$; $a=0.3, b=0.5$; and $a=0.3, b=0.7$. Smaller values of $a$ or larger values of $b$ lead to performance degradation. In other words, the choice of $a$ and $b$ needs to avoid values that are too small and too large, but also allow sufficient room for choice between $a$ and $b$. We believe this optimal range enables augmented data to incorporate richer variations while preserving fundamental preference knowledge from the original sequences. This enables the model to uncover accurate and deep-seated user sequential patterns, thereby enhancing the performance of tail recommendations. For $\alpha$, we find that optimal values tend to arise in $\{0.3,0.4,0.5\}$. When the value of $\alpha$ is small, the final augment representation is more likely to lean toward one of the representations involved in the mixing (either the original or the augmented) rather than having a balanced contribution from both. This uneven mixing operation reduces the quality of the final augmented representation.

\section{Conclusion}
This work introduces TADA, a tail-aware data augmentation for long-tail sequential recommendation. Our core idea is to tailor augmentation modules to increase interactions between tail items and users, thereby enhancing long-tail performance while avoiding the performance penalties caused by knowledge transfer \cite{liu2024fedbcgd,liuimproving}. To achieve that, we first capture the co-occurrence and correlation among low-popularity items by a linear model. Building upon this, we design two tail-aware augmentation operators, T-Substitute and T-Insert. The augmented and original sequences are mixed at the representation level to preserve preference knowledge. The cross augmentation further extends the mix operation across different tail-user sequences and augmented sequences to generate richer augmented samples. Extensive experiments show the effectiveness and generalizability of TADA. Our method can achieve a win-win situation for both head and tail, as well as overall performance. For future work, we hope to incorporate information from different modalities \cite{liu2023id,zhai2025heterogeneous,li2025aemvc} or utilize large language models \cite{liu2025cora,zhi2025reinventing,HILOWAGENT} to further enhance the accuracy and usability of our method.

%%
%% The acknowledgments section is defined using the "acks" environment
%% (and NOT an unnumbered section). This ensures the proper
%% identification of the section in the article metadata, and the
%% consistent spelling of the heading.
% \begin{acks}
% To Robert, for the bagels and explaining CMYK and color spaces.
% \end{acks}

%%
%% The next two lines define the bibliography style to be used, and
%% the bibliography file.

\bibliographystyle{ACM-Reference-Format}
\bibliography{references}

%%% -*-BibTeX-*-
%%% Do NOT edit. File created by BibTeX with style
%%% ACM-Reference-Format-Journals [18-Jan-2012].

\begin{thebibliography}{66}

%%% ====================================================================
%%% NOTE TO THE USER: you can override these defaults by providing
%%% customized versions of any of these macros before the \bibliography
%%% command.  Each of them MUST provide its own final punctuation,
%%% except for \shownote{} and \showURL{}.  The latter two
%%% do not use final punctuation, in order to avoid confusing it with
%%% the Web address.
%%%
%%% To suppress output of a particular field, define its macro to expand
%%% to an empty string, or better, \unskip, like this:
%%%
%%% \newcommand{\showURL}[1]{\unskip}   % LaTeX syntax
%%%
%%% \def \showURL #1{\unskip}           % plain TeX syntax
%%%
%%% ====================================================================

\ifx \showCODEN    \undefined \def \showCODEN     #1{\unskip}     \fi
\ifx \showISBNx    \undefined \def \showISBNx     #1{\unskip}     \fi
\ifx \showISBNxiii \undefined \def \showISBNxiii  #1{\unskip}     \fi
\ifx \showISSN     \undefined \def \showISSN      #1{\unskip}     \fi
\ifx \showLCCN     \undefined \def \showLCCN      #1{\unskip}     \fi
\ifx \shownote     \undefined \def \shownote      #1{#1}          \fi
\ifx \showarticletitle \undefined \def \showarticletitle #1{#1}   \fi
\ifx \showURL      \undefined \def \showURL       {\relax}        \fi
% The following commands are used for tagged output and should be
% invisible to TeX
\providecommand\bibfield[2]{#2}
\providecommand\bibinfo[2]{#2}
\providecommand\natexlab[1]{#1}
\providecommand\showeprint[2][]{arXiv:#2}

\bibitem[Bian et~al\mbox{.}(2022)]%
        {bian2022relevant}
\bibfield{author}{\bibinfo{person}{Shuqing Bian}, \bibinfo{person}{Wayne~Xin Zhao}, \bibinfo{person}{Jinpeng Wang}, {and} \bibinfo{person}{Ji-Rong Wen}.} \bibinfo{year}{2022}\natexlab{}.
\newblock \showarticletitle{A relevant and diverse retrieval-enhanced data augmentation framework for sequential recommendation}. In \bibinfo{booktitle}{\emph{CIKM}}. \bibinfo{pages}{2923--2932}.
\newblock


\bibitem[Box and Meyer(1986)]%
        {box1986analysis}
\bibfield{author}{\bibinfo{person}{George~EP Box} {and} \bibinfo{person}{R~Daniel Meyer}.} \bibinfo{year}{1986}\natexlab{}.
\newblock \showarticletitle{An analysis for unreplicated fractional factorials}.
\newblock \bibinfo{journal}{\emph{Technometrics}} \bibinfo{volume}{28}, \bibinfo{number}{1} (\bibinfo{year}{1986}), \bibinfo{pages}{11--18}.
\newblock


\bibitem[Chen et~al\mbox{.}(2022)]%
        {chen2022intent}
\bibfield{author}{\bibinfo{person}{Yongjun Chen}, \bibinfo{person}{Zhiwei Liu}, \bibinfo{person}{Jia Li}, \bibinfo{person}{Julian McAuley}, {and} \bibinfo{person}{Caiming Xiong}.} \bibinfo{year}{2022}\natexlab{}.
\newblock \showarticletitle{Intent contrastive learning for sequential recommendation}. In \bibinfo{booktitle}{\emph{WWW}}. \bibinfo{pages}{2172--2182}.
\newblock


\bibitem[Choi et~al\mbox{.}(2021)]%
        {choi2021session}
\bibfield{author}{\bibinfo{person}{Minjin Choi}, \bibinfo{person}{Jinhong Kim}, \bibinfo{person}{Joonseok Lee}, \bibinfo{person}{Hyunjung Shim}, {and} \bibinfo{person}{Jongwuk Lee}.} \bibinfo{year}{2021}\natexlab{}.
\newblock \showarticletitle{Session-aware linear item-item models for session-based recommendation}. In \bibinfo{booktitle}{\emph{WWW}}. \bibinfo{pages}{2186--2197}.
\newblock


\bibitem[Choi et~al\mbox{.}(2025)]%
        {choi2025linear}
\bibfield{author}{\bibinfo{person}{Minjin Choi}, \bibinfo{person}{Sunkyung Lee}, \bibinfo{person}{Seongmin Park}, {and} \bibinfo{person}{Jongwuk Lee}.} \bibinfo{year}{2025}\natexlab{}.
\newblock \showarticletitle{Linear Item-Item Models with Neural Knowledge for Session-based Recommendation}. In \bibinfo{booktitle}{\emph{SIGIR}}. \bibinfo{pages}{1666--1675}.
\newblock


\bibitem[Chuang~Zhao et~al\mbox{.}(2025)]%
        {HILOWAGENT}
\bibfield{author}{\bibinfo{person}{Hui~Tang Chuang~Zhao} {et~al\mbox{.}}} \bibinfo{year}{2025}\natexlab{}.
\newblock \bibinfo{title}{Grounded by Experience: Generative Healthcare Prediction Augmented with Hierarchical Agentic Retrieval}.
\newblock
\showeprint[arxiv]{2511.13293}


\bibitem[Dang et~al\mbox{.}(2024a)]%
        {dang2024repeated}
\bibfield{author}{\bibinfo{person}{Yizhou Dang}, \bibinfo{person}{Yuting Liu}, \bibinfo{person}{Enneng Yang}, \bibinfo{person}{Guibing Guo}, \bibinfo{person}{Linying Jiang}, \bibinfo{person}{Xingwei Wang}, {and} \bibinfo{person}{Jianzhe Zhao}.} \bibinfo{year}{2024}\natexlab{a}.
\newblock \showarticletitle{Repeated Padding for Sequential Recommendation}. In \bibinfo{booktitle}{\emph{RecSys}}. \bibinfo{pages}{497--506}.
\newblock


\bibitem[Dang et~al\mbox{.}(2025a)]%
        {dang2025data}
\bibfield{author}{\bibinfo{person}{Yizhou Dang}, \bibinfo{person}{Yuting Liu}, \bibinfo{person}{Enneng Yang}, \bibinfo{person}{Minhan Huang}, \bibinfo{person}{Guibing Guo}, \bibinfo{person}{Jianzhe Zhao}, {and} \bibinfo{person}{Xingwei Wang}.} \bibinfo{year}{2025}\natexlab{a}.
\newblock \showarticletitle{Data augmentation as free lunch: Exploring the test-time augmentation for sequential recommendation}. In \bibinfo{booktitle}{\emph{SIGIR}}. \bibinfo{pages}{1466--1475}.
\newblock


\bibitem[Dang et~al\mbox{.}(2023a)]%
        {dang2023ticoserec}
\bibfield{author}{\bibinfo{person}{Yizhou Dang}, \bibinfo{person}{Enneng Yang}, \bibinfo{person}{Guibing Guo}, \bibinfo{person}{Linying Jiang}, \bibinfo{person}{Xingwei Wang}, \bibinfo{person}{Xiaoxiao Xu}, \bibinfo{person}{Qinghui Sun}, {and} \bibinfo{person}{Hong Liu}.} \bibinfo{year}{2023}\natexlab{a}.
\newblock \showarticletitle{TiCoSeRec: Augmenting data to uniform sequences by time intervals for effective recommendation}.
\newblock \bibinfo{journal}{\emph{TKDE}} (\bibinfo{year}{2023}).
\newblock


\bibitem[Dang et~al\mbox{.}(2023b)]%
        {dang2023uniform}
\bibfield{author}{\bibinfo{person}{Yizhou Dang}, \bibinfo{person}{Enneng Yang}, \bibinfo{person}{Guibing Guo}, \bibinfo{person}{Linying Jiang}, \bibinfo{person}{Xingwei Wang}, \bibinfo{person}{Xiaoxiao Xu}, \bibinfo{person}{Qinghui Sun}, {and} \bibinfo{person}{Hong Liu}.} \bibinfo{year}{2023}\natexlab{b}.
\newblock \showarticletitle{Uniform sequence better: Time interval aware data augmentation for sequential recommendation}. In \bibinfo{booktitle}{\emph{AAAI}}.
\newblock


\bibitem[Dang et~al\mbox{.}(2024b)]%
        {dang2024data}
\bibfield{author}{\bibinfo{person}{Yizhou Dang}, \bibinfo{person}{Enneng Yang}, \bibinfo{person}{Yuting Liu}, \bibinfo{person}{Guibing Guo}, \bibinfo{person}{Linying Jiang}, \bibinfo{person}{Jianzhe Zhao}, {and} \bibinfo{person}{Xingwei Wang}.} \bibinfo{year}{2024}\natexlab{b}.
\newblock \showarticletitle{Data Augmentation for Sequential Recommendation: A Survey}.
\newblock \bibinfo{journal}{\emph{arXiv preprint arXiv:2409.13545}} (\bibinfo{year}{2024}).
\newblock


\bibitem[Dang et~al\mbox{.}(2025b)]%
        {dang2025augmenting}
\bibfield{author}{\bibinfo{person}{Yizhou Dang}, \bibinfo{person}{Jiahui Zhang}, \bibinfo{person}{Yuting Liu}, \bibinfo{person}{Enneng Yang}, \bibinfo{person}{Yuliang Liang}, \bibinfo{person}{Guibing Guo}, \bibinfo{person}{Jianzhe Zhao}, {and} \bibinfo{person}{Xingwei Wang}.} \bibinfo{year}{2025}\natexlab{b}.
\newblock \showarticletitle{Augmenting Sequential Recommendation with Balanced Relevance and Diversity}. In \bibinfo{booktitle}{\emph{AAAI}}, Vol.~\bibinfo{volume}{39}. \bibinfo{pages}{11563--11571}.
\newblock


\bibitem[Di et~al\mbox{.}(2025a)]%
        {di2025pifgsr}
\bibfield{author}{\bibinfo{person}{Yicheng Di}, \bibinfo{person}{Hongjian Shi}, \bibinfo{person}{Khushal~Haider Syed}, \bibinfo{person}{Ruhui Ma}, \bibinfo{person}{Haibing Guan}, \bibinfo{person}{Yuan Liu}, {and} \bibinfo{person}{Rajkumar Buyya}.} \bibinfo{year}{2025}\natexlab{a}.
\newblock \showarticletitle{PIFGSR: Pluggable framework for information fusion using generative artificial intelligence (GenAI) in recommender systems}.
\newblock \bibinfo{journal}{\emph{Information Fusion}} (\bibinfo{year}{2025}), \bibinfo{pages}{104004}.
\newblock


\bibitem[Di et~al\mbox{.}(2025b)]%
        {di2025federated}
\bibfield{author}{\bibinfo{person}{Yicheng Di}, \bibinfo{person}{Hongjian Shi}, \bibinfo{person}{Xiaoming Wang}, \bibinfo{person}{Ruhui Ma}, {and} \bibinfo{person}{Yuan Liu}.} \bibinfo{year}{2025}\natexlab{b}.
\newblock \showarticletitle{Federated recommender system based on diffusion augmentation and guided denoising}.
\newblock \bibinfo{journal}{\emph{TOIS}} \bibinfo{volume}{43}, \bibinfo{number}{2} (\bibinfo{year}{2025}), \bibinfo{pages}{1--36}.
\newblock


\bibitem[Di et~al\mbox{.}(2025c)]%
        {di2025personalized}
\bibfield{author}{\bibinfo{person}{Yicheng Di}, \bibinfo{person}{Xiaoming Wang}, \bibinfo{person}{Hongjian Shi}, \bibinfo{person}{Chongsheng Fan}, \bibinfo{person}{Rong Zhou}, \bibinfo{person}{Ruhui Ma}, {and} \bibinfo{person}{Yuan Liu}.} \bibinfo{year}{2025}\natexlab{c}.
\newblock \showarticletitle{Personalized consumer federated recommender system using fine-grained transformation and hybrid information sharing}.
\newblock \bibinfo{journal}{\emph{IEEE Transactions on Consumer Electronics}} (\bibinfo{year}{2025}).
\newblock


\bibitem[Fan et~al\mbox{.}(2021)]%
        {fan2021lighter}
\bibfield{author}{\bibinfo{person}{Xinyan Fan}, \bibinfo{person}{Zheng Liu}, \bibinfo{person}{Jianxun Lian}, \bibinfo{person}{Wayne~Xin Zhao}, \bibinfo{person}{Xing Xie}, {and} \bibinfo{person}{Ji-Rong Wen}.} \bibinfo{year}{2021}\natexlab{}.
\newblock \showarticletitle{Lighter and better: low-rank decomposed self-attention networks for next-item recommendation}. In \bibinfo{booktitle}{\emph{SIGIR}}. \bibinfo{pages}{1733--1737}.
\newblock


\bibitem[Fan et~al\mbox{.}(2022)]%
        {fan2022sequential}
\bibfield{author}{\bibinfo{person}{Ziwei Fan}, \bibinfo{person}{Zhiwei Liu}, \bibinfo{person}{Yu Wang}, \bibinfo{person}{Alice Wang}, \bibinfo{person}{Zahra Nazari}, \bibinfo{person}{Lei Zheng}, \bibinfo{person}{Hao Peng}, {and} \bibinfo{person}{Philip~S Yu}.} \bibinfo{year}{2022}\natexlab{}.
\newblock \showarticletitle{Sequential recommendation via stochastic self-attention}. In \bibinfo{booktitle}{\emph{WWW}}. \bibinfo{pages}{2036--2047}.
\newblock


\bibitem[He and McAuley(2016)]%
        {he2016fusing}
\bibfield{author}{\bibinfo{person}{Ruining He} {and} \bibinfo{person}{Julian McAuley}.} \bibinfo{year}{2016}\natexlab{}.
\newblock \showarticletitle{Fusing similarity models with markov chains for sparse sequential recommendation}. In \bibinfo{booktitle}{\emph{ICDM}}. IEEE, \bibinfo{pages}{191--200}.
\newblock


\bibitem[Hidasi et~al\mbox{.}(2015)]%
        {hidasi2015session}
\bibfield{author}{\bibinfo{person}{Bal{\'a}zs Hidasi}, \bibinfo{person}{Alexandros Karatzoglou}, \bibinfo{person}{Linas Baltrunas}, {and} \bibinfo{person}{Domonkos Tikk}.} \bibinfo{year}{2015}\natexlab{}.
\newblock \showarticletitle{Session-based recommendations with recurrent neural networks}.
\newblock \bibinfo{journal}{\emph{arXiv preprint arXiv:1511.06939}} (\bibinfo{year}{2015}).
\newblock


\bibitem[Hu et~al\mbox{.}(2022)]%
        {hu2022memory}
\bibfield{author}{\bibinfo{person}{Yidan Hu}, \bibinfo{person}{Yong Liu}, \bibinfo{person}{Chunyan Miao}, {and} \bibinfo{person}{Yuan Miao}.} \bibinfo{year}{2022}\natexlab{}.
\newblock \showarticletitle{Memory bank augmented long-tail sequential recommendation}. In \bibinfo{booktitle}{\emph{CIKM}}. \bibinfo{pages}{791--801}.
\newblock


\bibitem[Jang et~al\mbox{.}(2020)]%
        {jang2020cities}
\bibfield{author}{\bibinfo{person}{Seongwon Jang}, \bibinfo{person}{Hoyeop Lee}, \bibinfo{person}{Hyunsouk Cho}, {and} \bibinfo{person}{Sehee Chung}.} \bibinfo{year}{2020}\natexlab{}.
\newblock \showarticletitle{Cities: Contextual inference of tail-item embeddings for sequential recommendation}. In \bibinfo{booktitle}{\emph{ICMD}}. IEEE, \bibinfo{pages}{202--211}.
\newblock


\bibitem[Jiang et~al\mbox{.}(2021)]%
        {jiang2021sequential}
\bibfield{author}{\bibinfo{person}{Juyong Jiang}, \bibinfo{person}{Yingtao Luo}, \bibinfo{person}{Jae~Boum Kim}, \bibinfo{person}{Kai Zhang}, {and} \bibinfo{person}{Sunghun Kim}.} \bibinfo{year}{2021}\natexlab{}.
\newblock \showarticletitle{Sequential recommendation with bidirectional chronological augmentation of transformer}.
\newblock \bibinfo{journal}{\emph{arXiv preprint arXiv:2112.06460}}  \bibinfo{volume}{90} (\bibinfo{year}{2021}).
\newblock


\bibitem[Kang and McAuley(2018)]%
        {kang2018self}
\bibfield{author}{\bibinfo{person}{Wang-Cheng Kang} {and} \bibinfo{person}{Julian McAuley}.} \bibinfo{year}{2018}\natexlab{}.
\newblock \showarticletitle{Self-attentive sequential recommendation}. In \bibinfo{booktitle}{\emph{ICDM}}. IEEE, \bibinfo{pages}{197--206}.
\newblock


\bibitem[Kim et~al\mbox{.}(2023)]%
        {kim2023melt}
\bibfield{author}{\bibinfo{person}{Kibum Kim}, \bibinfo{person}{Dongmin Hyun}, \bibinfo{person}{Sukwon Yun}, {and} \bibinfo{person}{Chanyoung Park}.} \bibinfo{year}{2023}\natexlab{}.
\newblock \showarticletitle{Melt: Mutual enhancement of long-tailed user and item for sequential recommendation}. In \bibinfo{booktitle}{\emph{SIGIR}}. \bibinfo{pages}{68--77}.
\newblock


\bibitem[Kim et~al\mbox{.}(2019)]%
        {kim2019sequential}
\bibfield{author}{\bibinfo{person}{Yejin Kim}, \bibinfo{person}{Kwangseob Kim}, \bibinfo{person}{Chanyoung Park}, {and} \bibinfo{person}{Hwanjo Yu}.} \bibinfo{year}{2019}\natexlab{}.
\newblock \showarticletitle{Sequential and Diverse Recommendation with Long Tail.}. In \bibinfo{booktitle}{\emph{IJCAI}}, Vol.~\bibinfo{volume}{19}. \bibinfo{pages}{2740--2746}.
\newblock


\bibitem[Kingma and Ba(2014)]%
        {2014Adam}
\bibfield{author}{\bibinfo{person}{Diederik~P Kingma} {and} \bibinfo{person}{Jimmy Ba}.} \bibinfo{year}{2014}\natexlab{}.
\newblock \showarticletitle{Adam: A method for stochastic optimization}.
\newblock \bibinfo{journal}{\emph{arXiv preprint arXiv:1412.6980}} (\bibinfo{year}{2014}).
\newblock


\bibitem[Krichene and Rendle(2020)]%
        {krichene2020sampled}
\bibfield{author}{\bibinfo{person}{Walid Krichene} {and} \bibinfo{person}{Steffen Rendle}.} \bibinfo{year}{2020}\natexlab{}.
\newblock \showarticletitle{On sampled metrics for item recommendation}. In \bibinfo{booktitle}{\emph{KDD}}. \bibinfo{pages}{1748--1757}.
\newblock


\bibitem[Li et~al\mbox{.}(2017)]%
        {li2017neural}
\bibfield{author}{\bibinfo{person}{Jing Li}, \bibinfo{person}{Pengjie Ren}, \bibinfo{person}{Zhumin Chen}, \bibinfo{person}{Zhaochun Ren}, \bibinfo{person}{Tao Lian}, {and} \bibinfo{person}{Jun Ma}.} \bibinfo{year}{2017}\natexlab{}.
\newblock \showarticletitle{Neural attentive session-based recommendation}. In \bibinfo{booktitle}{\emph{CIKM}}. \bibinfo{pages}{1419--1428}.
\newblock


\bibitem[Li et~al\mbox{.}(2020)]%
        {li2020time}
\bibfield{author}{\bibinfo{person}{Jiacheng Li}, \bibinfo{person}{Yujie Wang}, {and} \bibinfo{person}{Julian McAuley}.} \bibinfo{year}{2020}\natexlab{}.
\newblock \showarticletitle{Time interval aware self-attention for sequential recommendation}. In \bibinfo{booktitle}{\emph{WSDM}}. \bibinfo{pages}{322--330}.
\newblock


\bibitem[Li et~al\mbox{.}(2025b)]%
        {li2025aemvc}
\bibfield{author}{\bibinfo{person}{Pengyuan Li}, \bibinfo{person}{Man Liu}, \bibinfo{person}{Dongxia Chang}, \bibinfo{person}{Yiming Wang}, \bibinfo{person}{Zisen Kong}, {and} \bibinfo{person}{Yao Zhao}.} \bibinfo{year}{2025}\natexlab{b}.
\newblock \showarticletitle{AEMVC: Mitigate Imbalanced Embedding Space in Multi-view Clustering}. In \bibinfo{booktitle}{\emph{MM}}. \bibinfo{pages}{6461--6470}.
\newblock


\bibitem[Li et~al\mbox{.}(2025a)]%
        {li2025reembedding}
\bibfield{author}{\bibinfo{person}{Zihao Li}, \bibinfo{person}{Yakun Chen}, \bibinfo{person}{Tong Zhang}, {and} \bibinfo{person}{Xianzhi Wang}.} \bibinfo{year}{2025}\natexlab{a}.
\newblock \showarticletitle{Reembedding and Reweighting are Needed for Tail Item Sequential Recommendation}. In \bibinfo{booktitle}{\emph{WWW}}. \bibinfo{pages}{4925--4936}.
\newblock


\bibitem[Li et~al\mbox{.}(2025c)]%
        {li2025listwise}
\bibfield{author}{\bibinfo{person}{Zihao Li}, \bibinfo{person}{Chao Yang}, \bibinfo{person}{Tong Zhang}, \bibinfo{person}{Yakun Chen}, \bibinfo{person}{Xianzhi Wang}, \bibinfo{person}{Guandong Xu}, {and} \bibinfo{person}{Daoyi Dong}.} \bibinfo{year}{2025}\natexlab{c}.
\newblock \showarticletitle{Listwise Preference Alignment Optimization for Tail Item Recommendation}.
\newblock \bibinfo{journal}{\emph{arXiv preprint arXiv:2507.02255}} (\bibinfo{year}{2025}).
\newblock


\bibitem[Liu et~al\mbox{.}(2021b)]%
        {liu2021pay}
\bibfield{author}{\bibinfo{person}{Hanxiao Liu}, \bibinfo{person}{Zihang Dai}, \bibinfo{person}{David So}, {and} \bibinfo{person}{Quoc~V Le}.} \bibinfo{year}{2021}\natexlab{b}.
\newblock \showarticletitle{Pay attention to mlps}.
\newblock \bibinfo{journal}{\emph{NIPS}}  \bibinfo{volume}{34} (\bibinfo{year}{2021}), \bibinfo{pages}{9204--9215}.
\newblock


\bibitem[Liu et~al\mbox{.}(2025a)]%
        {liuimproving}
\bibfield{author}{\bibinfo{person}{Junkang Liu}, \bibinfo{person}{Yuanyuan Liu}, \bibinfo{person}{Fanhua Shang}, \bibinfo{person}{Hongying Liu}, \bibinfo{person}{Jin Liu}, {and} \bibinfo{person}{Wei Feng}.} \bibinfo{year}{2025}\natexlab{a}.
\newblock \showarticletitle{Improving Generalization in Federated Learning with Highly Heterogeneous Data via Momentum-Based Stochastic Controlled Weight Averaging}. In \bibinfo{booktitle}{\emph{ICML}}.
\newblock


\bibitem[Liu et~al\mbox{.}(2024a)]%
        {liu2024fedbcgd}
\bibfield{author}{\bibinfo{person}{Junkang Liu}, \bibinfo{person}{Fanhua Shang}, \bibinfo{person}{Yuanyuan Liu}, \bibinfo{person}{Hongying Liu}, \bibinfo{person}{Yuangang Li}, {and} \bibinfo{person}{YunXiang Gong}.} \bibinfo{year}{2024}\natexlab{a}.
\newblock \showarticletitle{Fedbcgd: Communication-efficient accelerated block coordinate gradient descent for federated learning}. In \bibinfo{booktitle}{\emph{MM}}. \bibinfo{pages}{2955--2963}.
\newblock


\bibitem[Liu et~al\mbox{.}(2025b)]%
        {liu2025consistency}
\bibfield{author}{\bibinfo{person}{Junkang Liu}, \bibinfo{person}{Fanhua Shang}, \bibinfo{person}{Yuxuan Tian}, \bibinfo{person}{Hongying Liu}, {and} \bibinfo{person}{Yuanyuan Liu}.} \bibinfo{year}{2025}\natexlab{b}.
\newblock \showarticletitle{Consistency of local and global flatness for federated learning}. In \bibinfo{booktitle}{\emph{MM}}. \bibinfo{pages}{3875--3883}.
\newblock


\bibitem[Liu et~al\mbox{.}(2024b)]%
        {liu2024llm}
\bibfield{author}{\bibinfo{person}{Qidong Liu}, \bibinfo{person}{Xian Wu}, \bibinfo{person}{Yejing Wang}, \bibinfo{person}{Zijian Zhang}, \bibinfo{person}{Feng Tian}, \bibinfo{person}{Yefeng Zheng}, {and} \bibinfo{person}{Xiangyu Zhao}.} \bibinfo{year}{2024}\natexlab{b}.
\newblock \showarticletitle{Llm-esr: Large language models enhancement for long-tailed sequential recommendation}.
\newblock \bibinfo{journal}{\emph{NIPS}}  \bibinfo{volume}{37} (\bibinfo{year}{2024}), \bibinfo{pages}{26701--26727}.
\newblock


\bibitem[Liu et~al\mbox{.}(2023a)]%
        {liu2023diffusion}
\bibfield{author}{\bibinfo{person}{Qidong Liu}, \bibinfo{person}{Fan Yan}, \bibinfo{person}{Xiangyu Zhao}, \bibinfo{person}{Zhaocheng Du}, \bibinfo{person}{Huifeng Guo}, \bibinfo{person}{Ruiming Tang}, {and} \bibinfo{person}{Feng Tian}.} \bibinfo{year}{2023}\natexlab{a}.
\newblock \showarticletitle{Diffusion augmentation for sequential recommendation}. In \bibinfo{booktitle}{\emph{CIKM}}. \bibinfo{pages}{1576--1586}.
\newblock


\bibitem[Liu and Zheng(2020)]%
        {liu2020long}
\bibfield{author}{\bibinfo{person}{Siyi Liu} {and} \bibinfo{person}{Yujia Zheng}.} \bibinfo{year}{2020}\natexlab{}.
\newblock \showarticletitle{Long-tail session-based recommendation}. In \bibinfo{booktitle}{\emph{RecSys}}. \bibinfo{pages}{509--514}.
\newblock


\bibitem[Liu et~al\mbox{.}(2023b)]%
        {liu2023id}
\bibfield{author}{\bibinfo{person}{Yuting Liu}, \bibinfo{person}{Enneng Yang}, \bibinfo{person}{Yizhou Dang}, \bibinfo{person}{Guibing Guo}, \bibinfo{person}{Qiang Liu}, \bibinfo{person}{Yuliang Liang}, \bibinfo{person}{Linying Jiang}, {and} \bibinfo{person}{Xingwei Wang}.} \bibinfo{year}{2023}\natexlab{b}.
\newblock \showarticletitle{Id embedding as subtle features of content and structure for multimodal recommendation}.
\newblock \bibinfo{journal}{\emph{arXiv preprint arXiv:2311.05956}} (\bibinfo{year}{2023}).
\newblock


\bibitem[Liu et~al\mbox{.}(2025c)]%
        {liu2025cora}
\bibfield{author}{\bibinfo{person}{Yuting Liu}, \bibinfo{person}{Jinghao Zhang}, \bibinfo{person}{Yizhou Dang}, \bibinfo{person}{Yuliang Liang}, \bibinfo{person}{Qiang Liu}, \bibinfo{person}{Guibing Guo}, \bibinfo{person}{Jianzhe Zhao}, {and} \bibinfo{person}{Xingwei Wang}.} \bibinfo{year}{2025}\natexlab{c}.
\newblock \showarticletitle{Cora: Collaborative information perception by large language model’s weights for recommendation}. In \bibinfo{booktitle}{\emph{AAAI}}, Vol.~\bibinfo{volume}{39}. \bibinfo{pages}{12246--12254}.
\newblock


\bibitem[Liu et~al\mbox{.}(2021a)]%
        {liu2021contrastive}
\bibfield{author}{\bibinfo{person}{Zhiwei Liu}, \bibinfo{person}{Yongjun Chen}, \bibinfo{person}{Jia Li}, \bibinfo{person}{Philip~S Yu}, \bibinfo{person}{Julian McAuley}, {and} \bibinfo{person}{Caiming Xiong}.} \bibinfo{year}{2021}\natexlab{a}.
\newblock \showarticletitle{Contrastive self-supervised sequential recommendation with robust augmentation}.
\newblock \bibinfo{journal}{\emph{arXiv preprint arXiv:2108.06479}} (\bibinfo{year}{2021}).
\newblock


\bibitem[Liu et~al\mbox{.}(2021c)]%
        {liu2021augmenting}
\bibfield{author}{\bibinfo{person}{Zhiwei Liu}, \bibinfo{person}{Ziwei Fan}, \bibinfo{person}{Yu Wang}, {and} \bibinfo{person}{Philip~S Yu}.} \bibinfo{year}{2021}\natexlab{c}.
\newblock \showarticletitle{Augmenting sequential recommendation with pseudo-prior items via reversely pre-training transformer}. In \bibinfo{booktitle}{\emph{SIGIR}}. \bibinfo{pages}{1608--1612}.
\newblock


\bibitem[Liu et~al\mbox{.}(2022)]%
        {liu2022multi}
\bibfield{author}{\bibinfo{person}{Zhuang Liu}, \bibinfo{person}{Yunpu Ma}, \bibinfo{person}{Matthias Schubert}, \bibinfo{person}{Yuanxin Ouyang}, {and} \bibinfo{person}{Zhang Xiong}.} \bibinfo{year}{2022}\natexlab{}.
\newblock \showarticletitle{Multi-modal contrastive pre-training for recommendation}. In \bibinfo{booktitle}{\emph{ICMR}}. \bibinfo{pages}{99--108}.
\newblock


\bibitem[McAuley et~al\mbox{.}(2015)]%
        {mcauley2015inferring}
\bibfield{author}{\bibinfo{person}{Julian McAuley}, \bibinfo{person}{Rahul Pandey}, {and} \bibinfo{person}{Jure Leskovec}.} \bibinfo{year}{2015}\natexlab{}.
\newblock \showarticletitle{Inferring networks of substitutable and complementary products}. In \bibinfo{booktitle}{\emph{KDD}}. \bibinfo{pages}{785--794}.
\newblock


\bibitem[Qiu et~al\mbox{.}(2021)]%
        {qiu2021u}
\bibfield{author}{\bibinfo{person}{Zhaopeng Qiu}, \bibinfo{person}{Xian Wu}, \bibinfo{person}{Jingyue Gao}, {and} \bibinfo{person}{Wei Fan}.} \bibinfo{year}{2021}\natexlab{}.
\newblock \showarticletitle{U-BERT: Pre-training user representations for improved recommendation}. In \bibinfo{booktitle}{\emph{AAAI}}, Vol.~\bibinfo{volume}{35}. \bibinfo{pages}{4320--4327}.
\newblock


\bibitem[Qu et~al\mbox{.}(2022)]%
        {2022cmnrec}
\bibfield{author}{\bibinfo{person}{Shilin Qu}, \bibinfo{person}{Fajie Yuan}, \bibinfo{person}{Guibing Guo}, \bibinfo{person}{Liguang Zhang}, {and} \bibinfo{person}{Wei Wei}.} \bibinfo{year}{2022}\natexlab{}.
\newblock \showarticletitle{CmnRec: Sequential Recommendations with Chunk-accelerated Memory Network}.
\newblock \bibinfo{journal}{\emph{TKDE}} (\bibinfo{year}{2022}).
\newblock


\bibitem[Tan et~al\mbox{.}(2025)]%
        {tan2025taming}
\bibfield{author}{\bibinfo{person}{Shiyin Tan}, \bibinfo{person}{Dongyuan Li}, \bibinfo{person}{Renhe Jiang}, \bibinfo{person}{Zhen Wang}, \bibinfo{person}{Xingtong Yu}, {and} \bibinfo{person}{Manabu Okumura}.} \bibinfo{year}{2025}\natexlab{}.
\newblock \showarticletitle{Taming Recommendation Bias with Causal Intervention on Evolving Personal Popularity}. In \bibinfo{booktitle}{\emph{KDD}}. \bibinfo{pages}{2790--2800}.
\newblock


\bibitem[Tan et~al\mbox{.}(2016)]%
        {tan2016improved}
\bibfield{author}{\bibinfo{person}{Yong~Kiam Tan}, \bibinfo{person}{Xinxing Xu}, {and} \bibinfo{person}{Yong Liu}.} \bibinfo{year}{2016}\natexlab{}.
\newblock \showarticletitle{Improved recurrent neural networks for session-based recommendations}. In \bibinfo{booktitle}{\emph{DLRS}}. \bibinfo{pages}{17--22}.
\newblock


\bibitem[Tang and Wang(2018)]%
        {tang2018personalized}
\bibfield{author}{\bibinfo{person}{Jiaxi Tang} {and} \bibinfo{person}{Ke Wang}.} \bibinfo{year}{2018}\natexlab{}.
\newblock \showarticletitle{Personalized top-n sequential recommendation via convolutional sequence embedding}. In \bibinfo{booktitle}{\emph{WSDM}}. \bibinfo{pages}{565--573}.
\newblock


\bibitem[Tian et~al\mbox{.}(2023)]%
        {tian2023periodicity}
\bibfield{author}{\bibinfo{person}{Changxin Tian}, \bibinfo{person}{Binbin Hu}, \bibinfo{person}{Wayne~Xin Zhao}, \bibinfo{person}{Zhiqiang Zhang}, {and} \bibinfo{person}{Jun Zhou}.} \bibinfo{year}{2023}\natexlab{}.
\newblock \showarticletitle{Periodicity may be emanative: Hierarchical contrastive learning for sequential recommendation}. In \bibinfo{booktitle}{\emph{CIKM}}. \bibinfo{pages}{2442--2451}.
\newblock


\bibitem[Vaswani(2017)]%
        {vaswani2017attention}
\bibfield{author}{\bibinfo{person}{A Vaswani}.} \bibinfo{year}{2017}\natexlab{}.
\newblock \showarticletitle{Attention is all you need}.
\newblock \bibinfo{journal}{\emph{NIPS}} (\bibinfo{year}{2017}).
\newblock


\bibitem[Verma et~al\mbox{.}(2019)]%
        {verma2019manifold}
\bibfield{author}{\bibinfo{person}{Vikas Verma}, \bibinfo{person}{Alex Lamb}, \bibinfo{person}{Christopher Beckham}, \bibinfo{person}{Amir Najafi}, \bibinfo{person}{Ioannis Mitliagkas}, \bibinfo{person}{David Lopez-Paz}, {and} \bibinfo{person}{Yoshua Bengio}.} \bibinfo{year}{2019}\natexlab{}.
\newblock \showarticletitle{Manifold mixup: Better representations by interpolating hidden states}. In \bibinfo{booktitle}{\emph{ICML}}. PMLR, \bibinfo{pages}{6438--6447}.
\newblock


\bibitem[Xie et~al\mbox{.}(2022)]%
        {xie2022contrastive}
\bibfield{author}{\bibinfo{person}{Xu Xie}, \bibinfo{person}{Fei Sun}, \bibinfo{person}{Zhaoyang Liu}, \bibinfo{person}{Shiwen Wu}, \bibinfo{person}{Jinyang Gao}, \bibinfo{person}{Jiandong Zhang}, \bibinfo{person}{Bolin Ding}, {and} \bibinfo{person}{Bin Cui}.} \bibinfo{year}{2022}\natexlab{}.
\newblock \showarticletitle{Contrastive learning for sequential recommendation}. In \bibinfo{booktitle}{\emph{ICDE}}. IEEE, \bibinfo{pages}{1259--1273}.
\newblock


\bibitem[Yang et~al\mbox{.}(2023)]%
        {yang2023loam}
\bibfield{author}{\bibinfo{person}{Heeyoon Yang}, \bibinfo{person}{YunSeok Choi}, \bibinfo{person}{Gahyung Kim}, {and} \bibinfo{person}{Jee-Hyong Lee}.} \bibinfo{year}{2023}\natexlab{}.
\newblock \showarticletitle{LOAM: improving long-tail session-based recommendation via niche walk augmentation and tail session mixup}. In \bibinfo{booktitle}{\emph{SIGIR}}. \bibinfo{pages}{527--536}.
\newblock


\bibitem[Yin et~al\mbox{.}(2024)]%
        {yin2024dataset}
\bibfield{author}{\bibinfo{person}{Mingjia Yin}, \bibinfo{person}{Hao Wang}, \bibinfo{person}{Wei Guo}, \bibinfo{person}{Yong Liu}, \bibinfo{person}{Suojuan Zhang}, \bibinfo{person}{Sirui Zhao}, \bibinfo{person}{Defu Lian}, {and} \bibinfo{person}{Enhong Chen}.} \bibinfo{year}{2024}\natexlab{}.
\newblock \showarticletitle{Dataset regeneration for sequential recommendation}. In \bibinfo{booktitle}{\emph{KDD}}. \bibinfo{pages}{3954--3965}.
\newblock


\bibitem[Yuan et~al\mbox{.}(2019)]%
        {yuan2019simple}
\bibfield{author}{\bibinfo{person}{Fajie Yuan}, \bibinfo{person}{Alexandros Karatzoglou}, \bibinfo{person}{Ioannis Arapakis}, \bibinfo{person}{Joemon~M Jose}, {and} \bibinfo{person}{Xiangnan He}.} \bibinfo{year}{2019}\natexlab{}.
\newblock \showarticletitle{A simple convolutional generative network for next item recommendation}. In \bibinfo{booktitle}{\emph{WSDM}}. \bibinfo{pages}{582--590}.
\newblock


\bibitem[Yue et~al\mbox{.}(2024)]%
        {yue2024linear}
\bibfield{author}{\bibinfo{person}{Zhenrui Yue}, \bibinfo{person}{Yueqi Wang}, \bibinfo{person}{Zhankui He}, \bibinfo{person}{Huimin Zeng}, \bibinfo{person}{Julian McAuley}, {and} \bibinfo{person}{Dong Wang}.} \bibinfo{year}{2024}\natexlab{}.
\newblock \showarticletitle{Linear recurrent units for sequential recommendation}. In \bibinfo{booktitle}{\emph{WSDM}}. \bibinfo{pages}{930--938}.
\newblock


\bibitem[Zhai et~al\mbox{.}(2025a)]%
        {zhai2025heterogeneous}
\bibfield{author}{\bibinfo{person}{Chenhao Zhai}, \bibinfo{person}{Chang Meng}, \bibinfo{person}{Xueliang Wang}, \bibinfo{person}{Shuchang Liu}, \bibinfo{person}{Xiaolong Hu}, \bibinfo{person}{Shisong Tang}, \bibinfo{person}{Xiaoqiang Feng}, {and} \bibinfo{person}{Xiu Li}.} \bibinfo{year}{2025}\natexlab{a}.
\newblock \showarticletitle{Heterogeneous Multi-treatment Uplift Modeling for Trade-off Optimization in Short-Video Recommendation}.
\newblock \bibinfo{journal}{\emph{arXiv preprint arXiv:2511.18997}} (\bibinfo{year}{2025}).
\newblock


\bibitem[Zhai et~al\mbox{.}(2025b)]%
        {zhai2025combinatorial}
\bibfield{author}{\bibinfo{person}{Chenhao Zhai}, \bibinfo{person}{Chang Meng}, \bibinfo{person}{Yu Yang}, \bibinfo{person}{Kexin Zhang}, \bibinfo{person}{Xuhao Zhao}, {and} \bibinfo{person}{Xiu Li}.} \bibinfo{year}{2025}\natexlab{b}.
\newblock \showarticletitle{Combinatorial Optimization Perspective based Framework for Multi-behavior Recommendation}.
\newblock \bibinfo{journal}{\emph{arXiv preprint arXiv:2502.02232}} (\bibinfo{year}{2025}).
\newblock


\bibitem[Zhang et~al\mbox{.}(2017)]%
        {zhang2017mixup}
\bibfield{author}{\bibinfo{person}{Hongyi Zhang}, \bibinfo{person}{Moustapha Cisse}, \bibinfo{person}{Yann~N Dauphin}, {and} \bibinfo{person}{David Lopez-Paz}.} \bibinfo{year}{2017}\natexlab{}.
\newblock \showarticletitle{mixup: Beyond empirical risk minimization}.
\newblock \bibinfo{journal}{\emph{arXiv preprint arXiv:1710.09412}} (\bibinfo{year}{2017}).
\newblock


\bibitem[Zhang et~al\mbox{.}(2024)]%
        {zhang2024ninerec}
\bibfield{author}{\bibinfo{person}{Jiaqi Zhang}, \bibinfo{person}{Yu Cheng}, \bibinfo{person}{Yongxin Ni}, \bibinfo{person}{Yunzhu Pan}, \bibinfo{person}{Zheng Yuan}, \bibinfo{person}{Junchen Fu}, \bibinfo{person}{Youhua Li}, \bibinfo{person}{Jie Wang}, {and} \bibinfo{person}{Fajie Yuan}.} \bibinfo{year}{2024}\natexlab{}.
\newblock \showarticletitle{Ninerec: A benchmark dataset suite for evaluating transferable recommendation}.
\newblock \bibinfo{journal}{\emph{TPAMI}} (\bibinfo{year}{2024}).
\newblock


\bibitem[Zhao et~al\mbox{.}(2023a)]%
        {zhao2023sequential}
\bibfield{author}{\bibinfo{person}{Chuang Zhao}, \bibinfo{person}{Xinyu Li}, \bibinfo{person}{Ming He}, \bibinfo{person}{Hongke Zhao}, {and} \bibinfo{person}{Jianping Fan}.} \bibinfo{year}{2023}\natexlab{a}.
\newblock \showarticletitle{Sequential Recommendation via an Adaptive Cross-domain Knowledge Decomposition}. In \bibinfo{booktitle}{\emph{CIKM}}. \bibinfo{pages}{3453--3463}.
\newblock


\bibitem[Zhao et~al\mbox{.}(2023b)]%
        {zhao2023cross}
\bibfield{author}{\bibinfo{person}{Chuang Zhao}, \bibinfo{person}{Hongke Zhao}, \bibinfo{person}{Ming He}, \bibinfo{person}{Jian Zhang}, {and} \bibinfo{person}{Jianping Fan}.} \bibinfo{year}{2023}\natexlab{b}.
\newblock \showarticletitle{Cross-domain recommendation via user interest alignment}. In \bibinfo{booktitle}{\emph{WWW}}. \bibinfo{pages}{887--896}.
\newblock


\bibitem[Zhi et~al\mbox{.}(2025)]%
        {zhi2025reinventing}
\bibfield{author}{\bibinfo{person}{Xiaoquan Zhi}, \bibinfo{person}{Hongke Zhao}, \bibinfo{person}{Likang Wu}, \bibinfo{person}{Chuang Zhao}, {and} \bibinfo{person}{Hengshu Zhu}.} \bibinfo{year}{2025}\natexlab{}.
\newblock \showarticletitle{Reinventing Clinical Dialogue: Agentic Paradigms for LLM Enabled Healthcare Communication}.
\newblock \bibinfo{journal}{\emph{arXiv preprint arXiv:2512.01453}} (\bibinfo{year}{2025}).
\newblock


\bibitem[Zhou et~al\mbox{.}(2022)]%
        {zhou2022filter}
\bibfield{author}{\bibinfo{person}{Kun Zhou}, \bibinfo{person}{Hui Yu}, \bibinfo{person}{Wayne~Xin Zhao}, {and} \bibinfo{person}{Ji-Rong Wen}.} \bibinfo{year}{2022}\natexlab{}.
\newblock \showarticletitle{Filter-enhanced MLP is all you need for sequential recommendation}. In \bibinfo{booktitle}{\emph{WWW}}. \bibinfo{pages}{2388--2399}.
\newblock


\end{thebibliography}

%%
%% If your work has an appendix, this is the place to put it.
% \appendix
% \section{Research Methods}

\newpage

\appendix

\begin{algorithm}[!t]
\SetKwInput{KwInput}{Input}                
\SetKwInput{KwOutput}{Output} 
\KwInput{Item Embedding Matrix $M_V$, Interaction Matrix $M_D$, Original User Sequence $s_u$, Augmentation Hyperparameters $a,b,\alpha,\beta$}

\KwOutput{Augmented Representation $H_u^{ao},H_u^{ac}$}

Constructing correlation candidate set $cr_v$ based on $M_D$.\\

Constructing co-occurrence candidate set $cc_v$.\\

The union of set $cr_v$ and set $cc_v$ yields set $c_v$.\\

\tcp{Model Training Begins.}
Selecting operators based on Eq. (\ref{eq:select}).\\
\eIf{T-Substitute}{
Sampling probability $p$ and augmentation indices. \\
Augmenting $s_u$ based on Eq. (\ref{eq:T-Substitute}) to obtain $s_u{\prime}$.
}{
\tcp{Using T-Insert.}
Sampling probability $p$ and augmentation indices.\\
Augmenting $s_u$ based on Eq. (\ref{eq:T-Insert-a}) and Eq. (\ref{eq:T-Insert-b}) to obtain $s_u{\prime},s_u^e$ .
}
Mixup operation based on Eq. (\ref{eq:mixing-a}) and Eq. (\ref{eq:mixing-b}) to obtain $H_u^{ao}$.\\

Classifying sequences based on $\beta$.\\

Mixup operation based on Eq. (\ref{eq:shuffle}) and Eq. (\ref{eq:cross-mixing}) to obtain $H_u^{ac}$.\\

\textbf{Return} Augmented Representation $H_u^{ao},H_u^{ac}$

\caption{Pseudocode of TADA.}
\label{alg:algorithm}
\end{algorithm}

\section{Pseudocode of TADA}

We provide the pseudocode of TADA in Algorithm \ref{alg:algorithm}. It first constructs the correlation candidate set $cr_v$ and the co-occurrence candidate set $cc_v$. The final candidate set $c_v$ is the union of $cr_v$ and $cc_v$. Given an original user sequence $s_u$, we select tail-aware operators based on Equation (\ref{eq:select}). The selected operator augments the sequence based on the corresponding candidate set. Afterwards, we mixup at the representation level to obtain the final augmented representation $H_u^{ao}$. The cross augmentation further extends this augmentation across different tail and augmented sequences to obtain $H_u^{ac}$. All augmented samples are utilized for model training.

\section{Details of Baselines and Datasets}
\textbf{Backbone SR Models.} To validate the generalizability of our proposed TADA, we first selected three representative backbones. These three models are based on recurrent neural networks (RNNs), Transformers \cite{vaswani2017attention}, and multilayer perceptrons (MLPs), respectively.

\begin{itemize}
    \item \textbf{GRU4Rec} \cite{hidasi2015session}: It leverages gated recurrent units to model behavioral patterns. It argues that by modeling the whole session, more accurate recommendations can be provided.
    \item \textbf{SASRec} \cite{kang2018self}: It is a representative Transformer-based model, which leverages the multi-head self-attention for the next item recommendation task. It models the entire user sequence (without any recurrent or convolutional operations).
    \item \textbf{FMLPRec} \cite{zhou2022filter}: It finds that the filtering algorithms used in digital signal processing are effective in removing noise from sequence data. Inspired by this finding, an all-MLP model with learnable filters for SR tasks is proposed.
\end{itemize}

In Section \ref{sec:more_backbones}, We select three additional backbone SR models:

\begin{itemize}
    \item \textbf{NARM} \cite{li2017neural}: It incorporates attention mechanism into RNN architectures. It proposes to capture both the user's sequential behavior and main purpose in the current session.
    \item \textbf{LightSANs} \cite{fan2021lighter}: It introduces the low-rank decomposed self-attention, which projects the user's historical items into a small constant number of latent interests and leverages item-to-interest interaction to generate the context-aware representation.
    \item \textbf{gMLP} \cite{liu2021pay}: It suggests an alternative to the multi-head self-attention layers. It adopts a cross-token MLP structure named Spatial Gating Unit to capture sequential information.
\end{itemize}

\noindent \textbf{Data Augmentation SR Methods.} Since our approach involves data augmentation, we selected three representative augmentation methods. We do not select methods based on contrastive learning \cite{xie2022contrastive,tian2023periodicity}, as these introduce additional learning objectives. In contrast, all augmentation data generated by our TADA is utilized for the main recommendation task. DiffASR \cite{liu2023diffusion} and ASReP \cite{liu2021augmenting} are not involved since they have limitations on the type of backbones. Meanwhile, RepPad has reported better performance than them \cite{dang2024repeated}. Therefore, we chose RepPad as the baseline.

\begin{itemize}
    \item \textbf{CMR} \cite{xie2022contrastive}: It proposes three augmentation operators with contrastive learning. Here, we only use the three operators, including Crop, Mask, and Reorder.
    \item \textbf{CMRSI} \cite{liu2021contrastive}: Based on CMR, this work proposes two informative operators. We employ five operators: Crop, Mask, Reorder, Substitute, and Insert.
    \item \textbf{RepPad} \cite{dang2024repeated}: discoveries that traditional zero-value padding resulted in significant waste of input space, so it proposes using the original sequence as padding content to enhance model performance and training efficiency.
\end{itemize}

\noindent \textbf{Long-Tail SR Methods.}  We selected three representative long-tail SR methods for comparison. To ensure a fair comparison, we do not include LLM-ESR \cite{liu2024llm} and R$^2$REC \cite{li2025reembedding}, as they utilize additional modalities, such as text and images, whereas our proposed TADA only takes the sequence item ID as input. 

\begin{itemize}
    \item \textbf{CITIES} \cite{jang2020cities}: It enhances the quality of the tail-item embeddings by training an embedding-inference function using multiple contextual head items.
    \item \textbf{LOAM} \cite{yang2023loam}: It proposes Niche Walk Augmentation to capture the intrinsic characteristics in sessions with global information, considering long-tail items and Tail Session Mixup to make better tail representations.
    \item \textbf{MELT} \cite{kim2023melt}: It consists of bilateral branches, each of which is responsible for alleviating the long-tailed user and item problems, where the branches mutually enhance each other along with the curriculum learning strategy in an end-to-end manner.
\end{itemize}

\begin{table}[!t]
  \centering
 \caption{Statistics of datasets. The `AL' denotes average length.}
 \vspace{-1em}
 \scalebox{0.9}{
    \begin{tabular}{c|ccccc}
    \toprule
    \textbf{Dataset} & \textbf{\# Users} & \textbf{\# Items} & \textbf{\# Interactions} & \textbf{\# AL} & \textbf{Sparsity} \\
    \midrule
    Toys  & 19,412 & 11,924 & 167,597 & 8.6 & 99.93\% \\
    Beauty & 22,363 & 12,101 & 198,502 & 8.9 & 99.93\% \\
    Sports & 35,598 & 18,357 & 296,337 & 8.3 & 99.95\% \\
    Yelp & 30,431 & 20,033 & 316,354 & 10.4 & 99.95\% \\
    Douyin & 20,398 & 8,299 & 139,834 & 6.9  & 99.92\% \\
    \bottomrule
    \end{tabular}}%
  \label{tab:datasets}%
  \vspace{-1em}
\end{table}%

\begin{table*}[!t]
  \centering
  \caption{Performance comparison on Yelp and Douyin datasets. The best performance is bolded and the second-best is underlined.}
  \vspace{-1em}
      \renewcommand\arraystretch{0.85}
        \setlength{\tabcolsep}{1.5mm}{
  \scalebox{0.70}{
    \begin{tabular}{cc|cccc|cc|cccc|cc|cccc}
    \toprule
    \multirow{2}[2]{*}{Dataset} & \multirow{2}[2]{*}{Method} & \multicolumn{4}{c|}{Overall}  & \multicolumn{2}{c|}{Head Item} & \multicolumn{4}{c|}{Tail Item} & \multicolumn{2}{c|}{Head User} & \multicolumn{4}{c}{Tail User} \\
          &       & H@10  & N@10  & H@20  & N@20  & H@10  & N@10  & H@10  & N@10  & H@20  & N@20  & H@10  & N@10  & H@10  & N@10  & H@20  & N@20 \\
    \midrule
    \multirow{24}[6]{*}{Yelp} & GRU4Rec & 0.0168  & 0.0076  & 0.0328  & 0.0116  & 0.0281  & 0.0124  & 0.0060  & 0.0030  & 0.0117  & 0.0044  & 0.0168  & 0.0083  & 0.0168  & 0.0074  & 0.0330  & 0.0114  \\
          & CMR   & 0.0189  & 0.0097  & 0.0356  & 0.0133  & 0.0366  & 0.0180  & 0.0063  & 0.0031  & 0.0120  & 0.0045  & 0.0188  & 0.0087  & 0.0190  & 0.0082  & 0.0355  & 0.0121  \\
          & CMRSI & 0.0199  & 0.0990  & 0.0374  & 0.0142  & 0.0387  & 0.0196  & 0.0069  & 0.0031  & 0.0131  & 0.0047  & 0.0194  & 0.0092  & 0.0229  & 0.0101  & 0.0411  & 0.0139  \\
          & RepPad & 0.0207  & 0.0102  & 0.0394  & 0.0148  & 0.0409  & 0.0203  & 0.0071  & 0.0032  & \underline{0.0135}  & \underline{0.0049}  & 0.0211  & 0.0105  & 0.0247  & 0.0110  & 0.0447  & 0.0154  \\
          & CITIES & 0.0185  & 0.0087  & 0.0346  & 0.0127  & 0.0318  & 0.0148  & 0.0059  & 0.0030  & 0.0101  & 0.0040  & 0.0207  & 0.0097  & 0.0180  & 0.0085  & 0.0343  & 0.0126  \\
          & LOAM  & 0.0086  & 0.0038  & 0.0167  & 0.0059  & 0.0099  & 0.0044  & \underline{0.0074}  & \underline{0.0033}  & 0.0132  & 0.0047  & 0.0125  & 0.0057  & 0.0077  & 0.0034  & 0.0153  & 0.0053  \\
          & MELT  & \underline{0.0251}  & \underline{0.0120}  & \underline{0.0465}  & \underline{0.0173}  & \textbf{0.0514} & \textbf{0.0245} & 0.0000  & 0.0000  & 0.0005  & 0.0001  & \underline{0.0217}  & \underline{0.0110}  & \underline{0.0259}  & \underline{0.0122}  & \underline{0.0471}  & \underline{0.0175}  \\
          \rowcolor{mygray}\cellcolor{white} & TADA  & \textbf{0.0279} & \textbf{0.0135} & \textbf{0.0499} & \textbf{0.0190} & \underline{0.0492}  & \underline{0.0238}  & \textbf{0.0076} & \textbf{0.0037} & \textbf{0.0148} & \textbf{0.0054} & \textbf{0.0294} & \textbf{0.0143} & \textbf{0.0276} & \textbf{0.0133} & \textbf{0.0500} & \textbf{0.0189} \\
\cmidrule{2-18}          & SASRec & 0.0262  & 0.0129  & 0.0446  & 0.0175  & 0.0474  & 0.0235  & 0.0060  & 0.0027  & 0.0117  & 0.0042  & 0.0199  & 0.0096  & 0.0278  & 0.0137  & 0.0467  & 0.0184  \\
          & CMR   & 0.0272  & 0.0134  & 0.0474  & 0.0182  & 0.0493  & 0.0267  & 0.0064  & 0.0029  & 0.0129  & 0.0045  & 0.0214  & 0.0101  & 0.0263  & 0.0134  & 0.0450  & 0.0179  \\
          & CMRSI & 0.0296  & 0.0148  & 0.0511  & 0.0201  & 0.0528  & 0.0279  & 0.0069  & 0.0032  & 0.0136  & 0.0051  & 0.0237  & 0.0109  & 0.0295  & 0.0144  & 0.0501  & 0.0196  \\
          & RepPad & 0.0320  & 0.0161  & 0.0529  & 0.0208  & 0.0559  & 0.0288  & 0.0093  & 0.0047  & \underline{0.0170}  & 0.0065  & \underline{0.0280}  & \underline{0.0139}  & 0.0323  & \underline{0.0163}  & 0.0555  & 0.0212  \\
          & CITIES & 0.0261  & 0.0130  & 0.0449  & 0.0177  & 0.0480  & 0.0240  & 0.0051  & 0.0025  & 0.0114  & 0.0040  & 0.0196  & 0.0096  & 0.0277  & 0.0138  & 0.0473  & 0.0187  \\
          & LOAM  & 0.0136  & 0.0065  & 0.0240  & 0.0091  & 0.0174  & 0.0081  & \underline{0.0099}  & \underline{0.0050}  & 0.0166  & \underline{0.0066}  & 0.0133  & 0.0059  & 0.0136  & 0.0066  & 0.0243  & 0.0093  \\
          & MELT  & \underline{0.0327}  & \underline{0.0163}  & \underline{0.0559}  & \underline{0.0221}  & \textbf{0.0669} & \textbf{0.0333} & 0.0001  & 0.0000  & 0.0006  & 0.0002  & 0.0271  & 0.0131  & \textbf{0.0341} & \textbf{0.0171} & \textbf{0.0584} & \textbf{0.0232} \\
          \rowcolor{mygray}\cellcolor{white} & TADA  & \textbf{0.0332} & \textbf{0.0169} & \textbf{0.0563} & \textbf{0.0227} & \underline{0.0571}  & \underline{0.0292}  & \textbf{0.0105} & \textbf{0.0052} & \textbf{0.0176} & \textbf{0.0070} & \textbf{0.0315} & \textbf{0.0164} & \underline{0.0336}  & \textbf{0.0171} & \underline{0.0567}  & \underline{0.0228}  \\
\cmidrule{2-18}          & FMLP-Rec & 0.0196  & 0.0091  & 0.0349  & 0.0129  & 0.0345  & 0.0185  & 0.0053  & 0.0028  & 0.0094  & 0.0038  & 0.0191  & 0.0084  & 0.0193  & 0.0093  & 0.0348  & 0.0131  \\
          & CMR   & 0.0214  & 0.0097  & 0.0376  & 0.0142  & 0.0383  & 0.0203  & 0.0056  & 0.0031  & 0.0099  & 0.0040  & 0.0215  & 0.0097  & 0.0220  & 0.0102  & 0.0383  & 0.0149  \\
          & CMRSI & 0.0235  & 0.0116  & 0.0395  & 0.0164  & 0.0426  & 0.0211  & 0.0062  & 0.0036  & 0.0105  & 0.0042  & 0.0237  & 0.0116  & 0.0255  & 0.0116  & \underline{0.0415}  & 0.0156  \\
          & RepPad & 0.0249  & 0.0127  & 0.0409  & 0.0170  & 0.0487  & 0.0249  & \underline{0.0067}  & \textbf{0.0039} & \underline{0.0114}  & \underline{0.0044}  & \underline{0.0256}  & \underline{0.0130}  & \underline{0.0288}  & 0.0137  & 0.0455  & 0.0178  \\
          & CITIES & 0.0185  & 0.0089  & 0.0314  & 0.0121  & 0.0320  & 0.0153  & 0.0056  & 0.0027  & 0.0104  & 0.0039  & 0.0168  & 0.0081  & 0.0189  & 0.0091  & 0.0318  & 0.0123  \\
          & LOAM  & 0.0088  & 0.0042  & 0.0174  & 0.0064  & 0.0114  & 0.0050  & 0.0064  & 0.0035  & \textbf{0.0119} & \textbf{0.0049} & 0.0097  & 0.0049  & 0.0086  & 0.0041  & 0.0176  & 0.0063  \\
          & MELT  & \underline{0.0265}  & \underline{0.0132}  & \underline{0.0448}  & \underline{0.0177}  & \underline{0.0536}  & \underline{0.0267}  & 0.0007  & 0.0003  & 0.0012  & 0.0004  & 0.0210  & 0.0103  & 0.0279  & \underline{0.0139}  & 0.0467  & \underline{0.0186}  \\
          \rowcolor{mygray}\cellcolor{white} & TADA  & \textbf{0.0305} & \textbf{0.0153} & \textbf{0.0496} & \textbf{0.0201} & \textbf{0.0549} & \textbf{0.0274} & \textbf{0.0071} & 0.0037  & \textbf{0.0119} & \textbf{0.0049} & \textbf{0.0299} & \textbf{0.0150} & \textbf{0.0306} & \textbf{0.0153} & \textbf{0.0497} & \textbf{0.0201} \\
    \midrule
    \multirow{24}[6]{*}{Douyin} & GRU4Rec & 0.1023  & 0.0558  & 0.1467  & 0.0670  & 0.4002  & 0.2248  & 0.0639  & 0.0340  & 0.0973  & 0.0424  & 0.1424  & 0.0778  & 0.0923  & 0.0503  & 0.1334  & 0.0607  \\
          & CMR   & 0.1125  & 0.0627  & 0.1549  & 0.0719  & 0.4235  & 0.2318  & 0.0719  & 0.0387  & 0.0106  & 0.0473  & 0.1514  & 0.0832  & 0.0986  & 0.0560  & 0.1436  & 0.0677  \\
          & CMRSI & 0.1204  & 0.0670  & 0.1670  & 0.0787  & 0.4337  & 0.2391  & 0.0789  & 0.0448  & 0.1123  & 0.0516  & 0.1618  & 0.0887  & 0.1087  & 0.0605  & 0.1520  & 0.0719  \\
          & RepPad & \underline{0.1216}  & \underline{0.0677}  & \underline{0.1682}  & \underline{0.0795}  & \underline{0.4444}  & \underline{0.2423}  & \underline{0.0800}  & \underline{0.0452}  & \underline{0.1154}  & \underline{0.0532}  & \underline{0.1643}  & \underline{0.0900}  & \underline{0.1109}  & \underline{0.0621}  & \underline{0.1549}  & \underline{0.0733}  \\
          & CITIES & 0.1064  & 0.0566  & 0.1460  & 0.0666  & 0.4199  & 0.2266  & 0.0660  & 0.0347  & 0.0962  & 0.0423  & 0.1429  & 0.0751  & 0.0973  & 0.0520  & 0.1355  & 0.0617  \\
          & LOAM  & 0.0865  & 0.0453  & 0.1297  & 0.0562  & 0.2860  & 0.1352  & 0.0608  & 0.0337  & 0.0923  & 0.0416  & 0.1267  & 0.0674  & 0.0764  & 0.0398  & 0.1172  & 0.0500  \\
          & MELT  & 0.0509  & 0.0250  & 0.0889  & 0.0346  & 0.3839  & 0.1950  & 0.0080  & 0.0031  & 0.0256  & 0.0075  & 0.0635  & 0.0311  & 0.0478  & 0.0235  & 0.0819  & 0.0321  \\
          \rowcolor{mygray}\cellcolor{white} & TADA  & \textbf{0.1317} & \textbf{0.0737} & \textbf{0.1768} & \textbf{0.0850} & \textbf{0.4641} & \textbf{0.2572} & \textbf{0.0889} & \textbf{0.0500} & \textbf{0.1234} & \textbf{0.0587} & \textbf{0.1782} & \textbf{0.0972} & \textbf{0.1201} & \textbf{0.0678} & \textbf{0.1630} & \textbf{0.0786} \\
\cmidrule{2-18}          & SASRec & 0.1269  & 0.0743  & 0.1699  & 0.0851  & 0.4311  & 0.2478  & 0.0877  & 0.0519  & 0.1210  & 0.0603  & 0.1468  & 0.0809  & 0.1219  & 0.0726  & 0.1620  & 0.0827  \\
          & CMR   & 0.1305  & 0.0750  & 0.1739  & 0.0862  & 0.4423  & 0.2496  & 0.0886  & 0.0527  & 0.1247  & 0.0615  & 0.1490  & 0.0768  & 0.1241  & 0.0740  & 0.1676  & 0.0844  \\
          & CMRSI & 0.1332  & 0.0759  & 0.1806  & 0.0879  & 0.4504  & 0.2432  & 0.0923  & 0.0544  & 0.1283  & 0.0634  & 0.1535  & 0.0787  & 0.1281  & 0.0752  & 0.1730  & 0.0865  \\
          & RepPad & \underline{0.1400}  & \underline{0.0816}  & \underline{0.1894}  & \underline{0.0926}  & \underline{0.4564}  & \underline{0.2521}  & \underline{0.0972}  & \underline{0.0582}  & \underline{0.1396}  & \underline{0.0679}  & \underline{0.1617}  & \underline{0.0860}  & \underline{0.1304}  & \underline{0.0795}  & \underline{0.1796}  & \underline{0.0903}  \\
          & CITIES & 0.1190  & 0.0688  & 0.1630  & 0.0798  & 0.4088  & 0.2237  & 0.0816  & 0.0488  & 0.1145  & 0.0570  & 0.1412  & 0.0772  & 0.1134  & 0.0667  & 0.1543  & 0.0769  \\
          & LOAM  & 0.1156  & 0.0681  & 0.1518  & 0.0773  & 0.3233  & 0.1673  & 0.0888  & 0.0553  & 0.1135  & 0.0615  & 0.1351  & 0.0722  & 0.1107  & 0.0670  & 0.1449  & 0.0757  \\
          & MELT  & 0.0426  & 0.0226  & 0.0681  & 0.0290  & 0.3405  & 0.1851  & 0.0042  & 0.0016  & 0.0121  & 0.0036  & 0.0505  & 0.0272  & 0.0406  & 0.0214  & 0.0650  & 0.0275  \\
          \rowcolor{mygray}\cellcolor{white} & TADA  & \textbf{0.1500} & \textbf{0.0889} & \textbf{0.2022} & \textbf{0.1021} & \textbf{0.4757} & \textbf{0.2707} & \textbf{0.1080} & \textbf{0.0655} & \textbf{0.1516} & \textbf{0.0764} & \textbf{0.1829} & \textbf{0.0997} & \textbf{0.1418} & \textbf{0.0862} & \textbf{0.1915} & \textbf{0.0987} \\
\cmidrule{2-18}          & FMLP-Rec & 0.1212  & 0.0717  & 0.1579  & 0.0809  & 0.4135  & 0.2372  & 0.0836  & 0.0503  & 0.1120  & 0.0575  & 0.1468  & 0.0799  & 0.1148  & 0.0696  & 0.1485  & 0.0781  \\
          & CMR   & 0.1259  & 0.0764  & 0.1629  & 0.0876  & 0.4199  & 0.2473  & 0.0879  & 0.0519  & 0.1184  & 0.0620  & 0.1544  & 0.0861  & 0.1154  & 0.0732  & 0.1523  & 0.0844  \\
          & CMRSI & 0.1236  & 0.0722  & 0.1603  & 0.0825  & 0.4176  & 0.2402  & 0.0844  & 0.0510  & 0.1146  & 0.0596  & 0.1492  & 0.0827  & 0.1169  & 0.0716  & 0.1492  & 0.0803  \\
          & RepPad & \underline{0.1276}  & \underline{0.0793}  & \underline{0.1634}  & \underline{0.0894}  & \underline{0.4223}  & \underline{0.2529}  & \underline{0.0908}  & 0.0522  & \underline{0.1197}  & \underline{0.0637}  & \underline{0.1568}  & \underline{0.0872}  & \underline{0.1207}  & \underline{0.0752}  & \underline{0.1517}  & \underline{0.0862}  \\
          & CITIES & 0.1203  & 0.0709  & 0.1627  & 0.0817  & 0.3954  & 0.2271  & 0.0848  & 0.0508  & 0.1164  & 0.0588  & 0.1461  & 0.0786  & 0.1138  & 0.0690  & 0.1533  & 0.0790  \\
          & LOAM  & 0.1119  & 0.0671  & 0.1443  & 0.0753  & 0.3091  & 0.1598  & 0.0864  & \underline{0.0551}  & 0.1085  & 0.0607  & 0.1319  & 0.0745  & 0.1069  & 0.0653  & 0.1360  & 0.0726  \\
          & MELT  & 0.0562  & 0.0299  & 0.0846  & 0.0370  & 0.3886  & 0.2150  & 0.0134  & 0.0061  & 0.0288  & 0.0099  & 0.0677  & 0.0373  & 0.0534  & 0.0281  & 0.0808  & 0.0350  \\
          \rowcolor{mygray}\cellcolor{white} & TADA  & \textbf{0.1343} & \textbf{0.0824} & \textbf{0.1745} & \textbf{0.0925} & \textbf{0.4315} & \textbf{0.2574} & \textbf{0.0960} & \textbf{0.0598} & \textbf{0.1283} & \textbf{0.0680} & \textbf{0.1652} & \textbf{0.0899} & \textbf{0.1266} & \textbf{0.0805} & \textbf{0.1636} & \textbf{0.0898} \\
    \bottomrule
    \end{tabular}}}%
  \label{tab:main_2}%
\end{table*}%

\begin{table*}[!t]
  \centering
  \caption{Performance comparison of different methods for item cold start scenarios. The best performance is bolded.}
   \vspace{-1em}
 \scalebox{0.9}{
    \begin{tabular}{c|cc|cc|cc}
    \toprule
    \multirow{2}[2]{*}{Method} & \multicolumn{2}{c|}{Toys (3030/11924,25.4\%)} & \multicolumn{2}{c|}{Beauty (2962/12101,24.5\%)} & \multicolumn{2}{c}{Sports (4667/18357,25.4\%)} \\
          & H@10 & N@10 & H@10 & N@10 & H@10 & N@10 \\
    \midrule
    SASRec & 0.0435  & 0.0268  & 0.0211  & 0.0128  & 0.0149  & 0.0091  \\
    CMR   & 0.0479  & 0.0293  & 0.0237  & 0.0139  & 0.0156  & 0.0094  \\
    CMRSI & 0.0503  & 0.0309  & 0.0269  & 0.0193  & 0.0162  & 0.0096  \\
    RepPad & 0.0547  & 0.0325  & 0.0318  & 0.0236  & 0.0165  & 0.0097  \\
    CITIES & 0.0136  & 0.0056  & 0.0200  & 0.0095  & 0.0060  & 0.0028  \\
    LOAM  & 0.0420  & 0.0277  & 0.0296  & 0.0192  & 0.0121  & 0.0062  \\
    MELT  & 0.0119  & 0.0055  & 0.0011  & 0.0005  & 0.0001  & 0.0000  \\
    TADA & \textbf{0.0628} & \textbf{0.0404} & \textbf{0.0353} & \textbf{0.0252} & \textbf{0.0181} & \textbf{0.0114} \\
    \bottomrule
    \end{tabular}}%
    \vspace{-1em}
  \label{tab:item_cold}%
\end{table*}%

\begin{table*}[!t]
  \centering
  \caption{Performance comparison of different methods for user cold start scenarios. The best performance is bolded.}
   \vspace{-1em}
 \scalebox{0.9}{
    \begin{tabular}{c|cccc|cccc|cccc}
    \toprule
    \multirow{3}[4]{*}{Method} & \multicolumn{4}{c|}{Toys}     & \multicolumn{4}{c|}{Beauty}   & \multicolumn{4}{c}{Sports} \\
\cmidrule{2-13}          & \multicolumn{2}{c}{K=1} & \multicolumn{2}{c|}{K=2} & \multicolumn{2}{c}{K=1} & \multicolumn{2}{c|}{K=2} & \multicolumn{2}{c}{K=1} & \multicolumn{2}{c}{K=2} \\
          & H@10 & N@10 & H@10 & N@10 & H@10 & N@10 & H@10 & N@10 & H@10 & N@10 & H@10 & N@10 \\
    \midrule
    SASRec & 0.0482  & 0.0257  & 0.0358  & 0.0191  & 0.0443  & 0.0221  & 0.0349  & 0.0177  & 0.0235  & 0.0125  & 0.0178  & 0.0092  \\
    CMR   & 0.0503  & 0.0264  & 0.0367  & 0.0196  & 0.0465  & 0.0229  & 0.0362  & 0.0186  & 0.0246  & 0.0129  & 0.0186  & 0.0095  \\
    CMRSI & 0.0523  & 0.0280  & 0.0386  & 0.0210  & 0.0480  & 0.0238  & 0.0357  & 0.0182  & 0.0255  & 0.0132  & 0.0183  & 0.0093  \\
    RepPad & 0.0547  & 0.0294  & 0.0395  & 0.0215  & 0.0510  & 0.0256  & 0.0376  & 0.0192  & 0.0270  & 0.0138  & 0.0196  & 0.0099  \\
    CITIES & 0.0436  & 0.0229  & 0.0329  & 0.0171  & 0.0429  & 0.0218  & 0.0346  & 0.0171  & 0.0219  & 0.0115  & 0.0174  & 0.0088  \\
    LOAM  & 0.0365  & 0.0204  & 0.0275  & 0.0147  & 0.0365  & 0.0204  & 0.0275  & 0.0147  & 0.0145  & 0.0076  & 0.0132  & 0.0066  \\
    MELT  & 0.0345  & 0.0172  & 0.0274  & 0.0137  & 0.0347  & 0.0168  & 0.0277  & 0.0140  & 0.0242  & 0.0122  & 0.0193  & 0.0094  \\
    TADA  & \textbf{0.0567} & \textbf{0.0329} & \textbf{0.0429} & \textbf{0.0247} & \textbf{0.0530} & \textbf{0.0283} & \textbf{0.0394} & \textbf{0.0207} & \textbf{0.0287} & \textbf{0.0149} & \textbf{0.0211} & \textbf{0.0107} \\
    \bottomrule
    \end{tabular}}%
  \label{tab:user_cold}%
\end{table*}%

\begin{table}[!t]
  \centering
  \caption{Diversity comparison of different methods. The best performance is bolded.}
     \vspace{-1em}
 \scalebox{0.9}{
    \begin{tabular}{c|c|c|c}
    \toprule
    Method & TCov@5 & Method & TCov@5 \\
    \midrule
    \multicolumn{4}{c}{Beauty} \\
    \midrule
    GRU4Rec & 0.5719  & FMLPRec & 0.6432  \\
    CMR   & 0.5333  & CMR   & 0.5726  \\
    CMRSI & 0.5529  & CMRSI & 0.6215  \\
    RepPad & 0.5805  & RepPad & 0.6627  \\
    CITIES & 0.4504  & CITIES & 0.2168  \\
    LOAM  & 0.4227  & LOAM  & 0.6483  \\
    MELT  & 0.0053  & MELT  & 0.0872  \\
    TADA  & \textbf{0.5871} & TADA  & \textbf{0.6777} \\
    \midrule
    \multicolumn{4}{c}{Sports} \\
    \midrule
    GRU4Rec & 0.5381  & FMLPRec & 0.6916  \\
    CMR   & 0.4732  & CMR   & 0.6533  \\
    CMRSI & 0.5005  & CMRSI & 0.6208  \\
    RepPad & 0.5597  & RepPad & 0.7025  \\
    CITIES & 0.4287  & CITIES & 0.5917  \\
    LOAM  & 0.4588  & LOAM  & 0.6666  \\
    MELT  & 0.0010  & MELT  & 0.0319  \\
    TADA  & \textbf{0.5772} & TADA  & \textbf{0.7189} \\
    \bottomrule
    \end{tabular}}%
    \vspace{-1em}
  \label{tab:TCov}%
\end{table}%

\noindent \textbf{Datasets.} In addition to the dataset used in the main text, we add two additional datasets from other scenarios here. \textbf{Toys}, \textbf{Beauty}, and \textbf{Sports} are obtained from the Amazon shopping platform \cite{mcauley2015inferring}, corresponding to the ``Toys and Games'', ``Beauty'', and ``Sports and Outdoors'' categories. \textbf{Douyin} \cite{zhang2024ninerec} includes short-video watching records. \textbf{Yelp}\footnote{https://www.yelp.com/dataset} is a business dataset. We use the transaction records after January 1st, 2019. We reduce the data by extracting the 5-core \cite{liu2021contrastive,dang2023uniform,tian2023periodicity}. The maximum sequence length is set to 50. The statistics are summarized in Table \ref{tab:datasets}.

\section{More Results and Analysis}

\subsection{Results on Yelp and Douyin}

The experimental results of our TADA and different baseline methods with various backbone SR models on Yelp and Douyin datasets are presented in Table \ref{tab:main_2}. Our approach continues to demonstrate effectiveness and generalizability.

\subsection{Cold-Start Performance}
We conducted experiments in both user and item cold-start scenarios to validate TADA's effectiveness. Although we employed a 5-core, the leave-one-out strategy results in a significant number of items within the training set having extremely low interaction counts. For item cold starts, we selected items whose interaction frequency falls in the $1\leq X \leq 4$. We also provided the proportion of these items within the item set. The results are provided in Table \ref{tab:item_cold}. For user cold starts, we randomly removed K interactions (K=1 or 2) from each user during training to simulate cold-start scenarios. The results are provided in Table \ref{tab:user_cold}. Our TADA continues to demonstrate superior performance in cold-start scenarios. 

\begin{table}[!t]
  \centering
  \caption{Efficiency Comparison. We have provided the time required for training. Time required for candidate set construction of TADA is provided in parentheses.}
     \vspace{-1em}
 \scalebox{0.9}{
    \begin{tabular}{c|c|c|c}
    \toprule
    Method & Minutes & Method & Minutes \\
    \midrule
    \multicolumn{4}{c}{Beauty} \\
    \midrule
    GRU4Rec & 12.3  & FMLP-Rec & 21.3 \\
    CMR   & 23.3  & CMR   & 39.7 \\
    CMRSI & 242.5 & CMRSI & 279.4 \\
    RepPad & 19.4  & RepPad & 44.5 \\
    CITIES & 18.1  & CITIES & 23.2 \\
    LOAM  & 17.5  & LOAM  & 30.7 \\
    MELT  & 17.6  & MELT  & 36.5 \\
    TADA  & 18.0 (0.7) & TADA  & 25.9 (0.7) \\
    \midrule
    \multicolumn{4}{c}{Sports} \\
    \midrule
    GRU4Rec & 19.9  & FMLP-Rec & 37.6 \\
    CMR   & 29.8  & CMR   & 58.2 \\
    CMRSI & 372.3 & CMRSI & 442.3 \\
    RepPad & 24.6  & RepPad & 69.7 \\
    CITIES & 30.8  & CITIES & 36.2 \\
    LOAM  & 33.5  & LOAM  & 48.6 \\
    MELT  & 34.0  & MELT  & 61.8 \\
    TADA  & 29.1 (1.7) & TADA  & 35.35 (1.7) \\
    \bottomrule
    \end{tabular}}%
    \vspace{-1em}
  \label{tab:efficiency}%
\end{table}%

\subsection{Other Metrics}

We added TCov@5 (Tail Coverage@K from LOAM \cite{yang2023loam}) to measure how many tail-items are in recommendation lists, which can evaluate the diversity of TADA's recommendations. Its definition is as follows:
\begin{equation}
\text{Tail\_Coverage@K} = \frac{\left| \bigcup_{u \in \mathcal{U}} L_K^{\text{tail}}(u) \right|}{|\mathcal{V}_{\text{tail}}|}
\end{equation}
where $L_K(u)$ denotes the top-K recommendation list generated for user $u$. Specifically, $L_K^{\text{tail}}(u)$ represents the subset of $L_K(u)$ that contains items belonging to the long-tail item set $\mathcal{V}_{\text{tail}}$. The results are presented in Table \ref{tab:TCov}. We can observe that TADA enhances performance while increasing the frequency of tail-items. Although other methods can improve the performance, their contribution to recommendation diversity remains limited.

\subsection{Efficiency Comparison}

We provide the time required for different baselines and our proposed TADA in Table \ref{tab:efficiency} (Time required for candidate set construction is provided in parentheses). TADA demonstrates advantages in efficiency. Given the performance gains delivered by TADA, it offers significant practical value.

\end{document}